\documentclass[structabstract]{aa}  
\usepackage[english]{babel}
\usepackage{graphicx}
\usepackage{longtable}
\usepackage{natbib}
\bibpunct{(}{)}{;}{a}{}{,}

\usepackage[varg]{txfonts}
\usepackage{color}

\newcommand{\ms}{m\,s$^{-1}$}
\newcommand{\cd}{d$^{-1}$}
\newcommand{\mhz}{$\mu$Hz}

\usepackage{amstext}
\usepackage[normalem]{ulem}

\begin{document}

\authorrunning{F.~Borsa et al.}
\titlerunning{Putting exoplanets in the stellar context: $\tau$ Boo A}
   \title{The GAPS Programme with HARPS-N at TNG
}
   \subtitle{VII. Putting exoplanets in the stellar context: magnetic activity and asteroseismology of \object{$\tau$ Bootis A}\thanks{Based on observations made with the Italian Telescopio Nazionale Galileo (TNG)
operated on the island of La Palma by the Fundacion Galileo Galilei of the INAF at the
Spanish Observatorio Roque de los Muchachos of the IAC in the frame of the program 
Global Architecture of the Planetary Systems (GAPS).}
\thanks{Table 1 is only available in electronic form
at the CDS via anonymous ftp to cdsarc.u-strasbg.fr (130.79.128.5)
or via http://cdsweb.u-strasbg.fr/cgi-bin/qcat?J/A+A/}
}

     \author{F.~Borsa\inst{1}, G.~Scandariato\inst{2},
   M.~Rainer\inst{1}, A.~Bignamini\inst{3}, A.~Maggio\inst{4}, E.~Poretti\inst{1}, A.~F.~Lanza\inst{2}, M.~P.~Di Mauro\inst{5}, 
   S.~Benatti\inst{6}, K.~Biazzo\inst{2}, A.~S.~Bonomo\inst{7}, M.~Damasso\inst{7}, M.~Esposito\inst{8,9}, R.~Gratton\inst{6},
L.~Affer\inst{4}, M.~Barbieri\inst{10}, C.~Boccato\inst{6}, R.~U.~Claudi\inst{6},  R.~Cosentino\inst{2,11}, E. Covino\inst{12},  S.~Desidera\inst{6}, A.~F.~M.~Fiorenzano\inst{11}, D.~Gandolfi\inst{2,13},  A.~Harutyunyan\inst{11}, J.~Maldonado\inst{4}, G.~Micela\inst{4}, P.~Molaro\inst{3}, E.~Molinari\inst{11,14}, I.~Pagano\inst{2}, I.~Pillitteri\inst{4,15}, G.~Piotto\inst{6,16},   E.~Shkolnik\inst{17}, R.~Silvotti\inst{7}, R.~Smareglia\inst{3},  J.~Southworth\inst{18}, A.~Sozzetti\inst{7}, B.~Stelzer\inst{4}     
}
\offprints{F.~Borsa\\ \email{francesco.borsa@brera.inaf.it}}

\institute{ INAF -- Osservatorio Astronomico di Brera, Via E. Bianchi 46, 23807 Merate (LC), Italy
\and
INAF -- Osservatorio Astrofisico di Catania, Via S.Sofia 78, 95123, Catania, Italy
\and
INAF -- Osservatorio Astronomico di Trieste, via Tiepolo 11, 34143 Trieste, Italy
\and
INAF -- Osservatorio Astronomico di Palermo, Piazza del Parlamento, 1, 90134, Palermo, Italy
\and
INAF-IAPS Istituto di Astrofisica e Planetologia Spaziali, Via del Fosso del Cavaliere 100, 00133, Roma, Italy
\and
INAF -- Osservatorio Astronomico di Padova, Vicolo dell'Osservatorio 5, 35122, Padova, Italy
\and
INAF -- Osservatorio Astrofisico di Torino, Via Osservatorio 20, 10025, Pino Torinese, Italy
\and
Instituto de Astrof{\'i}sica de Canarias, C/Via L{\'a}ctea S/N, 38200 La Laguna, Tenerife, Spain
\and
Departamento de Astrof{\'i}sica, Universidad de La Laguna, 38205 La Laguna, Tenerife, Spain
\and
Department of Physics, University of Atacama, Copayapu 485, Copiapo, Chile
\and
Fundaci{\'o}n Galileo Galilei - INAF, Rambla Jos{\'e} Ana Fernandez P{\'e}rez 7, 38712 Bre$\tilde{\rm n}$a Baja, TF - Spain
\and
INAF -- Osservatorio Astronomico di Capodimonte, Salita Moiariello 16, 80131, Napoli, Italy
\and
Landessternwarte K$\ddot{\rm o}$nigstuhl, Zentrum f$\ddot{\rm u}$r Astronomie der Universitat Heidelberg, K$\ddot{\rm o}$nigstuhl 12, 69117 Heidelberg, Germany
\and
INAF - IASF Milano, via Bassini 15, 20133 Milano, Italy
\and
Harvard-Smithsonian Center for Astrophysics, 60 Garden Street, Cambridge, MA, 02138, USA
\and
Dip. di Fisica e Astronomia Galileo Galilei -- Universit$\grave{\rm a}$ di Padova, Vicolo dell'Osservatorio 2, 35122, Padova, Italy
\and
Lowell Observatory, 1400 W. Mars Hill Road, Flagstaff, AZ, 86001, USA
\and
Astrophysics Group, Keele University, Staffordshire, ST5 5BG, UK
}

   \date{Received ; accepted }

 
  \abstract
   {}
   {We observed the \object{$\tau$ Boo} system with the HARPS-N spectrograph to test a new observational strategy aimed at jointly studying asteroseismology, the planetary orbit, and star-planet magnetic interaction.} 
   {We collected high-cadence observations on 11 nearly consecutive nights and for each night averaged the raw FITS files using a
dedicated software. In this way we obtained spectra with a high signal-to-noise
ratio, used to study the variation of the \ion{Ca}{ii} H\&K lines and to have radial velocity values free from stellar oscillations, without losing the oscillations information.
We developed a dedicated software to build a new custom mask that we used to refine the radial velocity determination
 with the HARPS-N pipeline and perform the spectroscopic analysis.
}
   {We updated the planetary ephemeris and showed the acceleration caused by the stellar binary companion.
   Our results on the stellar activity variation suggest the
presence of a high-latitude plage during the time span of our observations. The correlation between the chromospheric activity and the planetary orbital phase remains unclear.
   Solar-like oscillations are detected in the radial velocity time series: we estimated asteroseismic quantities and found
that they agree well with theoretical predictions. Our stellar model yields \text{an age of }\text{\uline{\uline{}}} $0.9\pm0.5$ Gyr for $\tau$ Boo and further constrains the value of the stellar mass to $1.38\pm0.05$ M$_\odot$.
   }
   {}

   \keywords{ Stars: individual: $\tau$ Boo -- planetary systems -- Asteroseismology  -- techniques: spectroscopic -- Stars: activity
               }

   \maketitle
%


\section{Introduction\label{sec:intro}}
While the number of confirmed exoplanets is rapidly increasing, we are improving our ability to characterize 
them 
by studying their parent stars and the star-planet tidal or magnetic interactions.
At the moment, the most interesting targets for characterization are transiting exoplanets, but the large population of 
non-transiting planets solicits new methods and observation strategies for this purpose.

With the intent of characterizing planetary systems with a spectro-photometric approach, we selected the well-known system $\tau$ Bootis A (HD~120136, F6V, V=4.49) as a test case
in the 
Global Architecture of Planetary Systems\footnote{http://www.oact.inaf.it/exoit/EXO-IT/Projects/Entries/2011/12/
27\_GAPS.html} \citep[GAPS, ][]{2013A&A...554A..28C} programme. GAPS is a joint effort of Italian researchers in collaboration with a few experts abroad, and the GAPS team manages a long-term observational program with the high-precision HARPS-N spectrograph\footnote{http://www.tng.iac.es/instruments/harps/} \citep{2012SPIE.8446E..1VC} at the Telescopio Nazionale Galileo (TNG). $\tau$ Boo A is included in the 
subprogram dedicated to the characterization of planetary systems through studies of the interactions between the planets and their central stars. $\tau$ Boo's brightness allows for high-resolution spectroscopy and asteroseismology 
with a limited investment of telescope time.

The bright F6V star has a faint M2V companion \citep[$\tau$ Boo B, separation 1.83'',][]{2014AJ....147...65D},         
forming a long-period binary system.
The primary component (hereafter $\tau$ Boo for the sake of uniformity with current literature) was first claimed to host a planet with a period of 3.312 days by \citet{1997ApJ...474L.115B}; they used the radial velocity (RV) method.
Thanks to its brightness, this system was employed in the past to develop new analysis techniques that have later been applied to other stars. 
Among the most important discoveries are the definition of upper limits on reflected starlight, which provides a maximum value for the planet's albedo \citep{1999Natur.402..751C,2010A&A...514A..23R}, the spectroscopic detection of CO absorption lines in the planet atmosphere, which permitted determining the inclination angle of the system and thus the exact mass of the planet \citep[M$_{\rm p}$=$5.95\pm0.28$ M$_{\rm Jup}$, ][]{2012Natur.486..502B,2012ApJ...753L..25R} and, very recently, the first detection of water vapor in the atmosphere of a non-transiting exoplanet \citep{2014ApJ...783L..29L}.

$\tau$ Boo was also considered in searches for effects of star-planet magnetic interaction (SPMI). A peculiar characteristic of this star is the optically variable modulation of the light curve that is probably due to photospheric spots that persisted at fixed longitudes for a few hundred days, as observed by the MOST satellite \citep{2008A&A...482..691W}. SPMI is a long-debated issue, with the best evidence coming from a modulation of chromospheric activity tracers phased with the planetary orbital period rather than with the stellar rotation period \citep{2005ApJ...622.1075S,2008ApJ...676..628S,2009A&A...505..339L,2012A&A...544A..23L}. 

Assessment of this behavior requires long-term monitoring of stars with close-in massive planets (hot Jupiters), very high signal--to--noise spectra (S/N$>300$ at $3950 \AA$), and adequate resolving power ($R \ge 80000$) to measure variability in the core of deep chromospheric lines such as the \ion{Ca}{ii} H\&K doublet. Detecting periodicities equal to the planetary period is crucial because variability can be due to intrinsic stellar activity, not related to the presence of hot Jupiters. On the other hand, the study of stellar activity is important as a source of noise in the search and characterization of new planets in extra-solar systems.

In the case of $\tau$ Boo, previous searches for SPMI in the \ion{Ca}{ii} H\&K lines were ambiguous \citep{2005ApJ...622.1075S}. One possible reason is that SPMI is due to magnetic stresses between the stellar and planetary magnetic fields, but this effect is very limited in $\tau$ Boo because the stellar rotation is known to be synchronized with the orbital motion of the planet \citep[e.g., ][]{2000ApJ...531..415H}. Nonetheless, this system remains an interesting target because the parent star shows evidence of magnetic cycles, with yearly polarity reversals \citep[e.g., ][]{2007MNRAS.374L..42C,2008MNRAS.385.1179D,2009MNRAS.398.1383F}.

 Asteroseismology of stars hosting exoplanets received a great boost
from the photometric time series collected by space missions. 
On the other hand, spectroscopic campaigns aimed at detecting
solar-like oscillations are very hard to organize because a long time baseline is required to resolve and identify the
excited modes. For instance,  \citet{18sco} were able to obtain a measure of the large
separation of 18~Sco by means of  2833 data points collected over 12 complete nights.
It is difficult to insert an asteroseismic program in
a large project such as GAPS, which covers a multiplicity of subprograms and goals
with limited telescope time. Moreover, the exposure times have to be very short
to monitor
the solar-like oscillations, and consequently, the asteroseismic targets have to be
bright stars. 
Therefore we used
 the scientific case of $\tau$~Boo as a pathfinder, applying the high-cadence observational strategy
for asteroseismic purposes and matching it with some exoplanetary goals
(search for additional planets and study of the star-planet interaction).
The paper is structured as follows: Sect.~\ref{sec:datareduction} presents the observations and describes the data reduction. Section~\ref{sec:spectralAnalysis} is devoted to the spectral analysis, with results presented in Sect.~\ref{StellarparametersSect} (stellar parameters), Sect.~\ref{OrbitalFitSect} (orbital parameters), Sect.~\ref{sec:stellaractivity} (stellar activity), Sect.~\ref{sec:seismology} (asteroseismology), and Sect.~\ref{sec:evolution} (evolutionary stage). Conclusions are presented in Sect.~\ref{sec:conclu}.


\section{Observations and data reduction\label{sec:datareduction}}
$\tau$ Boo was observed with HARPS-N 
on 11 nights between April 13 and May 8, 2013. Very good phase coverage of the orbital period of the planet was obtained.
A few more observations were made in April and July 2014 to characterize the observed long-term trend.
Simultaneous Th-Ar calibration was used to achieve high RV precision.
The complete set of observations is shown in Table~\ref{RV_parziale}.
When possible, we observed the fast-rotating, hot
star $\eta$~UMa (B3V, V=1.84, V$\sin i = 205$ k\ms) immediately afterwards; this was our tool to remove telluric lines.

\begin{table}[!ht]
\begin{center}
\caption{HARPS-N RV observations of $\tau$ Boo. This table is available in its entirety online at the CDS.}
\setlength{\tabcolsep}{3pt}
\begin{tabular}{cccc}
\hline
\hline
BJD$_{\rm UTC}$-2450000. &      RV (\ms)        & RV err (\ms) &  Bis. span (\ms)  \\
\hline
\hline
\noalign{\smallskip}
6396.517465     &       -16145.21        &      0.83    &       -104.71 \\
6396.518460     &       -16136.13        &      0.83    &       -114.74 \\
6396.519444     &       -16141.63        &      0.84    &       -109.08 \\
6396.520439     &       -16141.12        &      0.80    &       -102.26 \\
... & ... & ... & ...\\
\noalign{\smallskip}
\hline
\hline
\end{tabular}
\label{RV_parziale}
\end{center}
\end{table}

To study the star-planet interaction, we needed to reach a high signal--to--noise
ratio (S/N) near the \ion{Ca}{ii} H\&K lines. 
To match this requirement with a single exposure would cause the saturation of the rest of the spectrum, 
given the spectral energy distribution of $\tau$ Boo and the lower efficiency of HARPS-N in the blue region.
The need to avoid saturation was combined with the requirements of an asteroseismic feasibility study of $\tau$~Boo with HARPS-N, allowing us to make a synergy between different science themes of the GAPS campaign.
Our observational strategy consisted of taking several one-minute 
exposures and then averaging them to obtain a single spectrum with a very high S/N.
In this way, we obtained both the high-cadence spectra to monitor the solar-like oscillations and the mean high S/N spectra to 
study the \ion{Ca}{ii} H\&K variations over the orbital period of the planet. 
Moreover, we also obtained RV values to study the long-term variability of the system.

We developed an interactive program written in Python \citep{internalsum} to average the raw images pixel by pixel, adjusting the header of the created files.
The resulting mean FITS files are ready to be passed through the HARPS-N pipeline. In this way they are reduced 
exactly in the same way as all the FITS files acquired with HARPS-N (Table~\ref{RVmeanfiles}).

The Julian dates of the mean files calculated by the pipeline are not the correct dates because of the overhead time between the single exposures.This time is $\sim$25 sec, therefore its influence on consecutive 1 min exposures is relevant.
We corrected for this by taking an average of the Julian 
dates of the single images, weighted on their respective S/N: in this way, we introduced a sort of exposure-meter 
information in the mean file.
The RVs were then corrected for the change in the barycentric Earth radial velocity (BERV) between the pipeline-estimated Julian date and the corrected date.
For this purpose, we created a new tool and verified it to be comparable with the HARPS-N pipeline at the level of $\sim3$~c\ms\,\citep{internalsum}.

\begin{table}[!ht]
\begin{center}
\caption{RVs for the $\tau$~Boo nightly mean spectra. $\phi$ refers to the orbital phase of the planetary companion, based on the ephemeris of Sect.~\ref{OrbitalFitSect} and considering the planetary inferior conjunction as $\phi$=0.}
\setlength{\tabcolsep}{5pt}
\begin{tabular}{ccccc}
\hline
\hline
Night & BJD$_{\rm UTC}$-2450000 & RV [\ms] & RV err [\ms]  & $\phi$\\
\hline
\hline
\noalign{\smallskip}
1 & 6396.531788 &   -16135.08   &  0.86 &       0.73\\
2 & 6397.536059 &        -16716.81      & 0.77 &        0.03\\
3 & 6398.499820 &        -17016.77      & 0.93 &        0.33\\ 
4 & 6399.548081 &        -16231.25      & 1.39 &        0.65\\
5 & 6401.498724 &        -17085.13      & 0.79 &        0.23\\ 
6 & 6402.497328 &        -16501.11         &1.31 &      0.53\\
7 & 6406.653040 &        -16169.98      & 0.78 &        0.79\\
8 & 6407.703825 &        -16904.41      & 0.89 &        0.10\\
9 & 6408.686713 &        -16878.51      &  0.96 &       0.40\\
10 & 6410.680969 &       -16603.65      & 1.34 &        0.00\\
11 & 6421.517542 &       -17054.54        & 0.97 &      0.27\\
\noalign{\smallskip}
\hline
\hline
\end{tabular}
\label{RVmeanfiles}
\end{center}
\end{table}

The data (324 single exposures and 11 mean exposures) were finally
reduced using the HARPS-N data
reduction software (DRS) pipeline on the Yabi platform. 
Yabi \citep{YABI} is a Python web application installed at IA2\footnote{http://ia2.oats.inaf.it/} in Trieste that allows authorized users to run the HARPS-N DRS
pipeline on their own proprietary data with custom input parameters.

The pipeline installed at the TNG estimates the radial velocities of the targets by computing a cross-correlation function 
\citep[CCF, ][]{2002A&A...388..632P}
using the best-suited line mask of the available masks (G2, K5, or M2 spectral type). 
Taking advantage of the Yabi platform, we were able to create and implement a new custom mask for $\tau$~Boo \citep{internalmask,internalgratton}. Using the 
standard G2 mask as a starting point, we measured the depths of several unblended lines in a 
well-exposed, high S/N $\tau$~Boo spectrum. From these measurements we 
found the following empirical correlation between the line depths (LD) of the G2 mask and those of the 
$\tau$~Boo spectrum:
$LD_{\tau~Boo} = 0.0409321\cdot e^{2.80654\cdot LD_{G2}}$.
Using this equation, we corrected the values of the line depths in the standard G2 mask and created the new custom mask (3625 photospheric lines)
that we used for the data reduction. 
In addition to this, we increased the width of the
half-window of the weighted CCF from the default value of 20 up to 30~k\ms \,(about twice the value of V$\sin i$ of the star) to clearly cover the continuum around the wings of the mean line profile. 
These improvements reduced the RV errors by $\sim$5\%.


\section{Spectral analysis\label{sec:spectralAnalysis}}
The analysis of the magnetic activity of $\tau$ Boo was performed using the nightly high S/N mean spectra.
The CCF provided by the DRS pipeline was computed by cross-correlating the spectrum of $\tau$ Boo with the custom mask (see Sect.~\ref{sec:datareduction}): it is thus essentially a high S/N mean line profile, and it is highly sensitive to the presence of active regions on the stellar photosphere.

To study the variability of the photospheric activity, we monitored the line profile variations of the CCF, in particular its contrast (CCFc) and full-width at half-maximum (FWHM), the bisector inverse slope \citep[BIS,][]{Gray2008}, and the v$_{asy}$ parameter \citep{Figueira2013}. We computed these indicators by means of custom-made routines, developed by our team for HARPS-N spectra.

We also monitored the variability of the chromospheric activity. We focused on the \ion{Ca}{ii} H\&K lines (3968.47~\AA\ and 3933.66~\AA), the \ion{Na}{i} D$_{\rm 12}$ doublet (5889.95~\AA\ and 5895.92~\AA), the \ion{He}{i} D$_{\rm 3}$ triplet (blend at 5875.62~\AA), and the H$_{\rm\alpha}$ line (6562.79~\AA) as chromospheric diagnostics.

The chromospheric indicators were analyzed starting from the HARPS-N one-dimensional spectra reduced with the DRS pipeline, which computes the localization of spectral orders on the two-dimensional images, performs order extraction and stitching, corrects for flat-fielding, rejects cosmic-rays, and calibrates the wavelength. Our analysis is based on the differential comparison between the spectra in the series realigned in the wavelength space.
This approach allows us to avoid the uncertainties related to the absolute flux calibration and continuum normalization and is therefore more robust than the method adopted by \citet{Scandariato2013}, which depends on the normalization to the continuum (a difficult task in the presence of broad lines such as the \ion{Ca}{ii} H\&K and order stitching). 

Here we discuss how we reduced the data to perform the differential analysis, while in Sect.~\ref{sec:activity} we analyze the variability in the spectra of $\tau$ Boo.
In the following analysis we excluded the data point of night 4 because it has low S/N, and no spectroscopic standard star was observed because of bad weather.

The \ion{Na}{i} D$_{\rm 12}$ doublet, the \ion{He}{i} D$_{\rm 3}$ triplet, and the H$_{\rm\alpha}$ line are affected by telluric contamination, which is variable from night to night as a result
of different airmasses and sky conditions. For each night of observation, telluric features were removed from the target spectrum by comparing it with the spectrum of the standard star $\eta$~UMa. 
We corrected for telluric absorption using the task \texttt{telluric} in IRAF \citep{Tody1993}. In contrast, the \ion{Ca}{ii} H{\&}K spectral region of the standard star does not show any telluric feature above noise. Telluric correction was therefore not performed on the corresponding spectral interval of $\tau$ Boo to preserve the original S/N.

To carry out the spectral differential analysis, we needed to rescale all the spectra to the same flux scale. To this purpose, we selected the spectrum at mid-time (i.e.,\ the spectrum of night 6) with respect to which we performed the differential spectral analysis. This choice aims at minimizing any time-dependent instrumental effect along the time series of spectra (see below).

\begin{figure}
\centering
\includegraphics[viewport=1 1 340 235,clip,width=\linewidth]{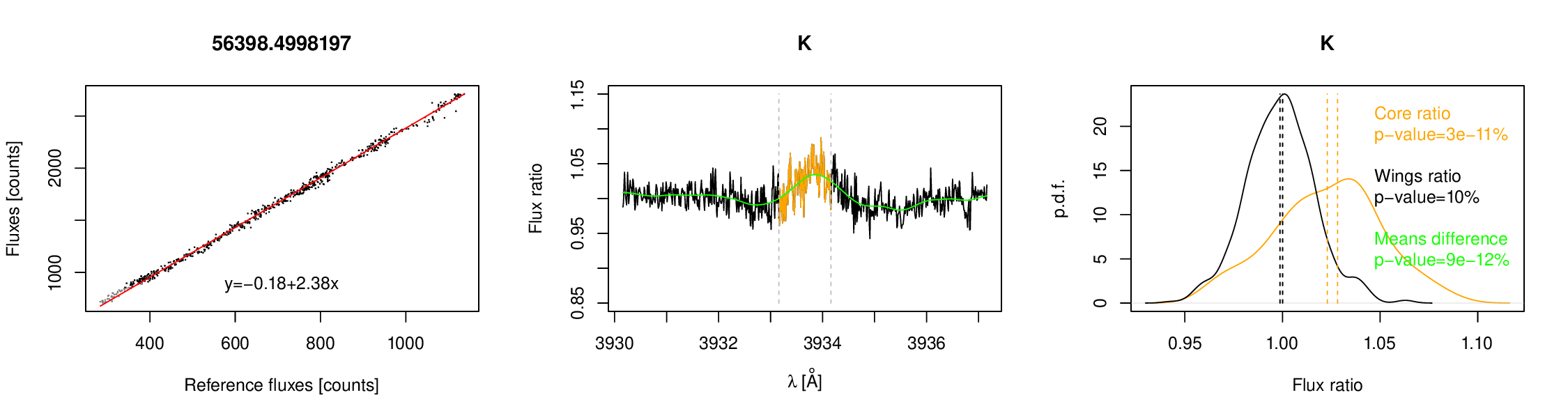}
\caption{\footnotesize{Example of the flux calibration of the mean spectrum of night 3 over the mean spectrum of night 6 in the \ion{Ca}{ii} K spectral range. The best fit (red line) is computed using the black dots, which represent instrumental fluxes out of the line core. The instrumental fluxes of the line core (in gray) are not included in the fit. The equation of the best fit is also shown.}\label{fig:comparison}}
\end{figure}

For each diagnostic we extracted the corresponding spectral region, and we compared each spectrum in the series with the reference spectrum on a pixel-by-pixel basis (Fig.~\ref{fig:comparison}), taking advantage of the wavelength stability of HARPS-N. In this way, the reference spectrum defined the common flux scale, and all the other spectra were rescaled accordingly using a linear best fit of the flux vs\ flux relation. The intercept adjusts the background level, while the slope rescales intrinsic stellar fluxes. Line cores were excluded from the best fit to avoid any intrinsic nonlinearity that might have been introduced by chromospheric variability. The widths of the analyzed spectral ranges and the line cores are reported in Table~\ref{tab:widths}.

\begin{table}
\begin{center}
\caption{Spectral ranges used for the analyzed chromospheric diagnostics.}\label{tab:widths}
\footnotesize
\begin{tabular}{ccccc}
 \hline\hline
Line & Vacuum wavelength& Spectral width & Core width\\
     & (\AA) & (\AA) & (\AA)\\
\hline\hline
\ion{Ca}{ii} K  & 3933.66 & 3.5 &  0.5\\
\ion{Ca}{ii} H  & 3968.47 & 3.5 &  0.5\\
\ion{Na}{i} D$_{\rm 2}$  & 5889.95 & 3.5 &  0.3\\
\ion{Na}{i} D$_{\rm 1}$  & 5895.92 & 3.5 &  0.3\\
\ion{He}{i} D$_{\rm 3}$  & 5875.72 & 3.5 &  0.6\\
H$_{\rm\alpha}$  & 6562.79 & 3.5 &  0.5\\
 \hline\hline
\end{tabular}
\end{center}
\end{table}

\begin{figure}
\centering
\includegraphics[viewport=0 0 340 240,clip,width=\linewidth]{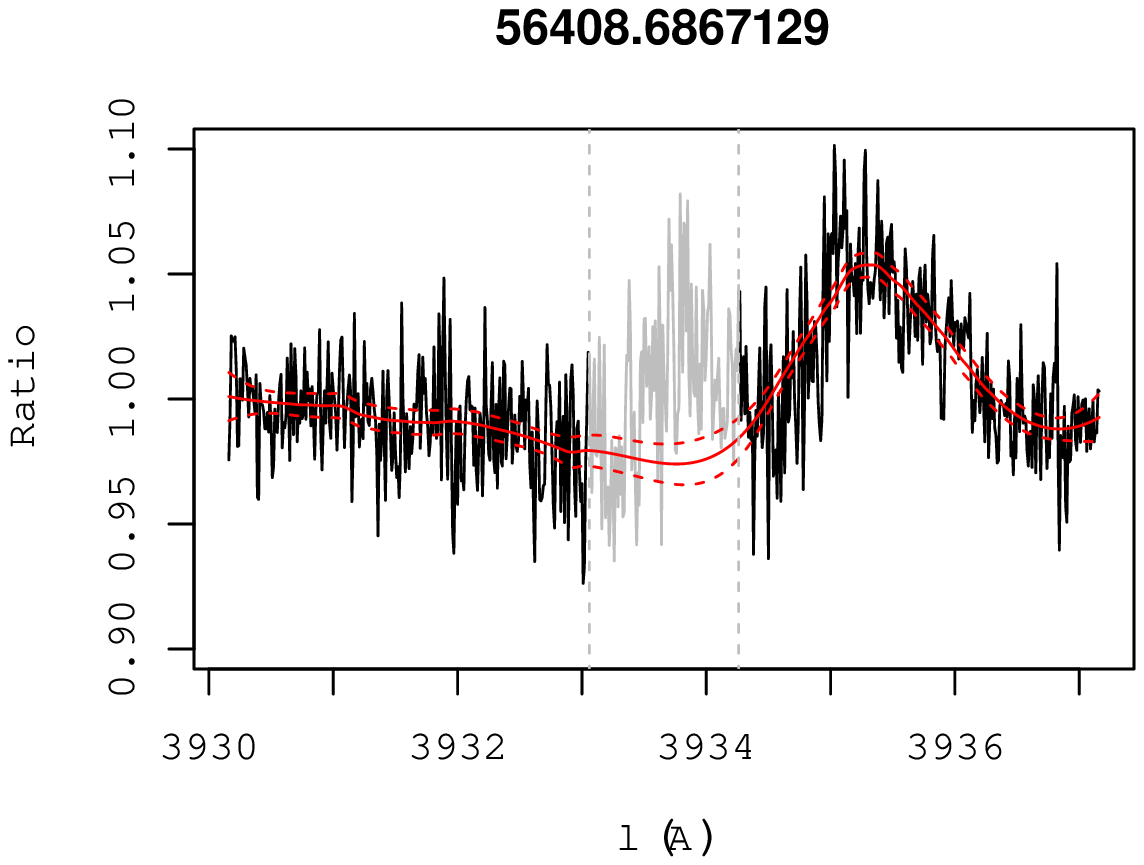}
\caption{\footnotesize{Example of the low-order correction of the \ion{Ca}{ii} K line. The black line is the ratio of one spectrum in the series to the reference spectrum, while the red line is the low-order fit (95\% confidence band is shown with red dashes). The vertical dashed lines bracket the line core, in gray, which is excluded from the fit.}\label{fig:low_order}}
\end{figure}

This procedure generally led to satisfactory results, that is,\ fluxes were well aligned along a straight line (Fig.~\ref{fig:comparison}). Still, in a few cases especially in the \ion{Ca}{ii} K line, the residuals of the fit were higher by $\sim$5\% than the noise
over small wavelength ranges, typically $\lesssim$1~\AA\ (Fig.~\ref{fig:low_order}).
These distortions, frequently observed during the first year of operation of HARPS-N, are not relevant in the measurement of RVs, but introduce spectral inhomogeneities on a night-to-night basis that may hamper the SPMI analysis. To correct for them, we divided the spectra pixel by pixel by the reference spectrum, and we locally fitted the ratio with a low-order polynomial, excluding a narrow window centered on the line cores where the SPMI signal might be mistaken as an instrumental effect. Finally, each spectrum was divided by the fitted ratio to remove the low-order flux variations. In Fig.~\ref{fig:low_order} we show an example of the low-order correction in the \ion{Ca}{ii} K line.

\begin{figure}[!ht]
\centering
\includegraphics[viewport=30 27 576 360,clip,width=\linewidth]{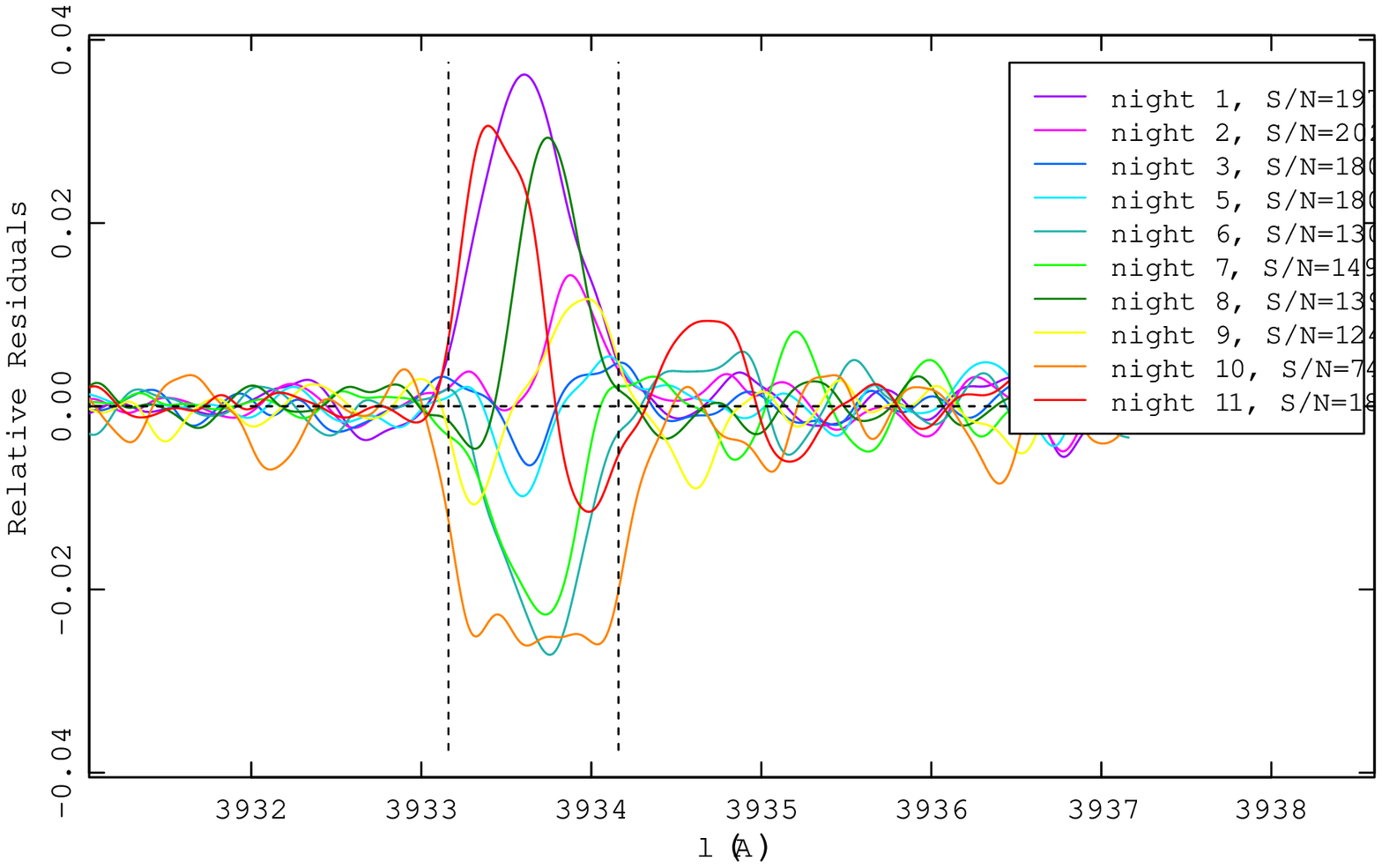}
\includegraphics[viewport=30 27 576 360,clip,width=\linewidth]{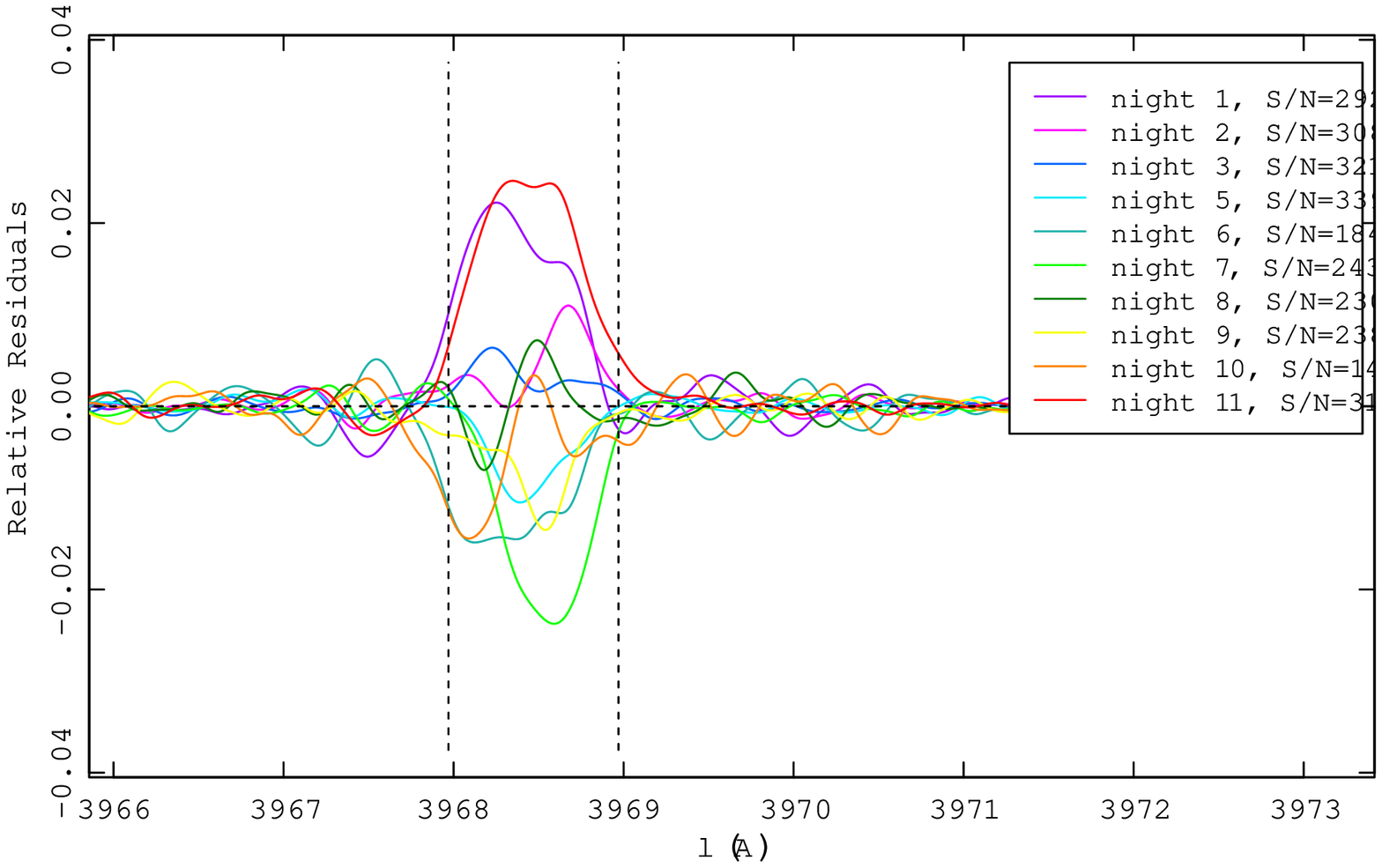}
\caption{\footnotesize{Residuals relative to the average spectrum of the \ion{Ca}{ii} K (top panel) and H (bottom panel) lines. The color code (online version only) is reported in the legend, together with the S/N in the line for a 1~\AA\ spectral element (night 4 has been excluded, see text). The residuals have been smoothed to avoid cluttering; for this reason, they slightly exceed the vertical dashed lines.}\label{fig:residuals}}
\end{figure}

After removing the distortions, we averaged the time series of the corrected mean spectra for each spectral range.
Then we computed the relative residuals for each mean spectrum with respect to the averaged spectrum 
and integrated them over a $\sim$1\AA\ window around the line center to obtain the integrated relative deviation (IRD).
Figure~\ref{fig:residuals} shows the residuals, which have been smoothed 
to show the variability in the \ion{Ca}{ii} H\&K line core more
clearly.
We computed the uncertainties on the IRDs by summing in quadrature the photon noise and the uncertainties on flux calibration and low-order correction. In this way, we take into account the same sources of error more than once, thus obtaining an upper limit to the true uncertainty.

We note that the relative residuals in the line cores are comparable to or slightly smaller than our low-order corrections to the spectra. The reliability of our approach rests on the verification that the recovered \ion{Ca}{ii} H\&K signals (IRDs) are reasonably correlated, and the same holds for \ion{Ca}{ii} H+K vs H$_{\rm\alpha}$ IRDs (Sect.~\ref{sec:activity}). We have verified that our low-order correction improves these physically meaningful correlations.


\section{Stellar parameters\label{StellarparametersSect}}

We derived the physical parameters of $\tau$~Boo by means of fitting synthetic spectra \citep[ATLAS9 stellar atmosphere models,][]{2004astro.ph..5087C} to a normalized HARPS-N mean file using the SME software by \citet{1996AA.118.595}. The hydrogen lines could not be used to determine the temperature because they span several spectral orders, which introduces problems in the normalization of the spectra. Therefore we used the region between 5160 and 5190 \AA, where the Mg triplet is found.

\begin{figure}
\begin{center}
\includegraphics[angle=270, width=8cm]{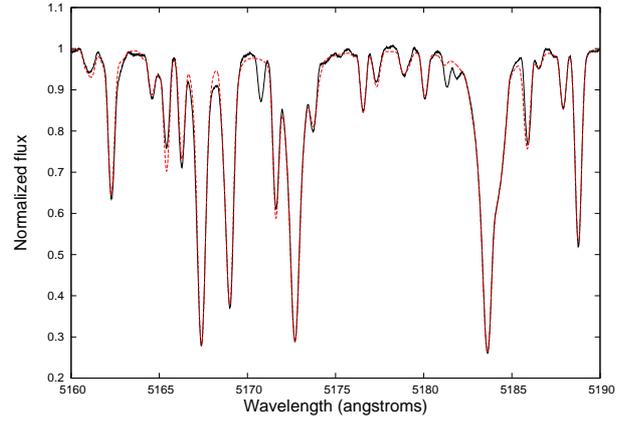}
\caption{\footnotesize{Fit of the observed mean spectrum (solid black line) with the synthetic spectrum (dashed red line) in the wavelength region considered, using the SME software.}}
\label{synt-spec}
\end{center}
\end{figure}

We computed the V$\sin{i}$ using the Fourier transform of the CCF of a mean spectrum (see Sect.~\ref{DiffRotSect}). 
This value was kept fixed in the fit while the other parameters were left free to vary.
Table~\ref{stellar_param} shows our results, which agree well with the values found in the literature \citep[e.g.,][]{2013A&A...556A.150S}.
A portion of the observed and synthetic spectra is shown in Fig.~\ref{synt-spec}.
An independent check of the stellar parameters was made with the method
based on the equivalent widths of spectral absorption lines \citep[][and references therein]{2007A&A...469..783S,2012MNRAS.427.2905B}.
Even if this method is not preferable because of the high V$\sin{i}$ value of $\tau$~Boo, results were compatible with those estimated with the synthetic spectra method. 

\begin{table}[!ht]
\begin{center}
\caption{Stellar parameters derived for $\tau$~Boo.}
\begin{tabular}{cc}
\hline
\hline
Parameter & Value \\
\hline
\hline
\noalign{\smallskip}
T$_{\mathrm{eff}}$ [K] & $6399\pm 45$ \\   

$\log g$ [c\ms] & $4.27\pm 0.06$ \\ 

[Fe/H] &   $0.26\pm 0.03$ \\  

V$\sin i$ [k\ms]  & $14.27 \pm 0.06$ \\  

Luminosity [$L_{\odot}$]  & $3.06\pm 0.16$   \\

Mass [$M_{\odot}$]  & $1.39\pm 0.25$ \\

Radius [$R_{\odot}$]  & $1.42\pm 0.08$ \\

V$_{\mathrm{breakup}}$ [k\ms]  & $352\pm 67$   \\
\noalign{\smallskip}
\hline
\hline
\end{tabular}
\label{stellar_param}
\end{center}
\end{table}

We determined stellar radius and mass (Table~\ref{stellar_param}) by combining the spectroscopic results with the parallax \citep[$64.03\pm0.19$ mas,][]{2007A&A...474..653V} and using a bolometric correction \citep[$-0.0078,$][]{2010AJ....140.1158T,1996ApJ...469..355F}.

\subsection{Differential rotation\label{DiffRotSect}}

We found evidence of solar-like differential rotation in $\tau$~Boo from studying the Fourier transform of the mean line profiles of the HARPS-N mean spectra. We used two different mean line profiles: the CCF computed by the HARPS-N pipeline and the profile obtained using the LSD software \citep{1997MNRAS.291.658D} on the wavelength regions 4415-4805, 4915-5285, and 5365-6505 {\AA}.
The CCF was computed using the Yabi interface, which means that we were able to reduce the spectra with our custom mask, and we further increased the CCF half-window width up to 40 k\ms, to ensure that we captured the whole mean line profile and the continuum.

We used a Fourier transform on both the LSD and the CCF profiles (Fig.~\ref{fourier}) and found the values of the first two zero positions $q_{1}$ and $q_{2}$. The $q_{1}$ position gives an estimate of the projected rotational velocity V$\sin{i}$, while the ratio $q_{2}/q_{1}$ is an indicator of differential rotation \citep{2002A&A...384..155R}, either solar-like ($q_{2}/q_{1} < 1.72$) or antisolar ($q_{2}/q_{1} > 1.83$). We found $q_{2}/q_{1} = 1.53 \pm 0.08$ for the LSD profile and $q_{2}/q_{1} = 1.63 \pm 0.04$ for the CCF. The larger error in the LSD profile arises probably from the fact that we did not use the whole spectral range. 
These values are compatible with the results from \citet{2006A&A...446..267R} ($q_{2}/q_{1} = 1.57 \pm 0.04$) and \citet{2007MNRAS.374L..42C} ($q_{2}/q_{1} = 1.60 \pm 0.02$).

The $q_{2}/q_{1}$ we found can indicate either solar-like differential rotation (equator rotating faster than the poles) or strong gravity-darkening in a rigidly rotating star. In the second case, the following empirical equation \citep{2003AA...408..707R} can be used to infer the rotational velocity of the star:

\begin{figure}
\begin{center}
\includegraphics[angle=270, width=8cm]{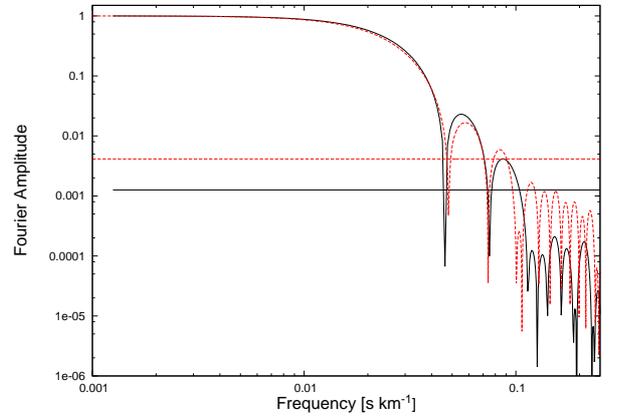}
\caption{\footnotesize{Amplitude of the Fourier transform of a mean line profile. 
Red dashed line refers to the LSD software, black solid line to the HARPS-N CCF.
The horizontal lines show the noise level.}}
\label{fourier}
\end{center}
\end{figure}

\begin{equation}
q_{2}/q_{1} = 1.75 + av + bv^{2}
,\end{equation}
where $a$ and $b$ are parameters that depend on the stellar spectral type. In particular, these parameters are $a = 0.172\times10^{-3}$ and $b = -0.993\times10^{-6}$ for F0 stars and $a=0.184\times10^{-3}$ and $b=-0.116\times10^{-5}$ for G0 stars. 
$\tau$~Boo is classified as an F6 star, falling in the middle of these values, which means that to justify the values of $q_{2}/q_{1}$ found by means of rigid rotation, the star's rotational velocity should be either between 415 and 450 k\ms (in the CCF case) or between 520 and 550 k\ms (in the LSD case). Considering the values of mass and radius 
of Table~\ref{stellar_param},
we can compute the breakup velocity $v_{\rm{breakup}} = 352$~k\ms. 

For the values $q_{2}/q_{1}$ to be caused by gravity-darkening, the star would have to rotate faster than the breakup velocity.
We can therefore reliably establish that the two $q_{2}/q_{1}$ values found show solar-like differential rotation in $\tau$~Boo.

Adopting an inclination angle for the star $i = 40^\circ$ \citep{2008MNRAS.385.1179D} and using the $q_{2}/q_{1}$ values, we computed the differential rotation parameter $\alpha = \frac{\Delta\Omega}{{\Omega}_{0}}$,where ${\Omega}_{0}$ is the equatorial angular velocity of the star and $\Delta\Omega$ is the difference between the equatorial and polar angular velocities.
Using the empirical relation \citep{2003A&A...398..647R}
\begin{equation}
\frac{\alpha}{\sqrt{\sin{i}}} = 2.74 - 5.16\left(\frac{q_2}{q_1}\right) + 4.32\left(\frac{q_2}{q_1}\right)^2 - 1.30\left(\frac{q_2}{q_1}\right)^3
,\end{equation}
we found $\alpha = 0.24 \pm 0.07$ from $q_{2}/q_{1}=1.53$ (LSD profile) and $\alpha = 0.16 \pm 0.04$ from $q_{2}/q_{1}=1.63$ (CCF),
both compatible with the value $\alpha = 0.18$ found by \citet{2007MNRAS.374L..42C}. The LSD result is exactly the same as the differential rotation found by \citet{2008MNRAS.385.1179D} using spectropolarimetric measures.

Our work shows that the LSD profile and the CCF provide similar values of $q_{2}/q_{1}$. As such, the CCF computed internally by the HARPS-N pipeline can be used as an indicator of differential rotation, at least for high S/N spectra.


\section{Orbital fit\label{OrbitalFitSect}}

We used data from Table~\ref{RV_parziale}
to perform the orbital fit, but when multiple exposures were taken during one night, we used the RV value corresponding to the nightly mean spectra (e.g., Table~\ref{RVmeanfiles}) to be independent of stellar oscillations. The final HARPS-N set of measurements is composed of 20 RV points.
In addition to our HARPS-N data, we used the recently released archival data of the Lick Observatory \citep{2014ApJS..210....5F}.
These data were obtained with four different iodine-cell setups (identified as number 2, 13, 6, and 8 in Table \ref{fitresults_tauboo}) that were taken in consideration during the analysis adding a RV shift as a free parameter to each set. We excluded all the data with $\sigma>30$ \ms (e.g., all the data of Lick setup number 2)
and those that did not maintain the same instrumental setup continuously. We used 166 Lick RV measurements in the orbital fit. 
By combining Lick and HARPS-N data, we obtained a dataset composed of 186 measurements, spanning a time interval of about 20 years.

\begin{figure}[!h]
\begin{center}
\includegraphics[trim = 0cm 0cm 0cm 0.8cm, clip, width=\linewidth]{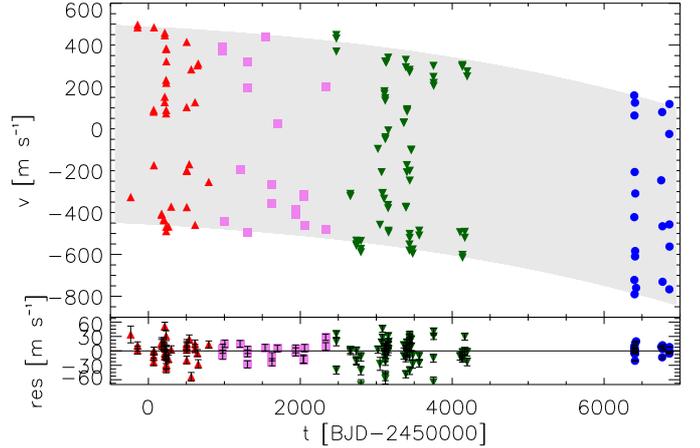}
\caption{\footnotesize{Two-objects orbital fit (gray) for the Lick Observatory archival data (red upward triangles, pink squares, and green downward triangles corresponding to the different setup configurations 13, 6, and 8 respectively) and HARPS-N data (blue circles). RV residuals are shown in the bottom plot.
}}
\label{AST01_fit}
\end{center}
\end{figure}

We fitted the RVs with a Levenberg-Marquardt algorithm \citep{2009ApJS..182..205W}, 
using a two-planet model (Fig.~\ref{AST01_fit}) to take into account both the short-term variability  due to the known planet and the long-term trend due to the stellar binary companion (Sect.~\ref{sect:binary}). 
To estimate the error-bars, we used a bootstrapping code \citep{2012ApJ...761...46W}. 
The orbital parameters obtained are reported in Table~\ref{fitresults_tauboo}.

\begin{table}[!ht]
\begin{center}
\caption{\footnotesize{Orbital parameters estimated for the $\tau$~Boo system. Period and periastron of the 
binary companion $\tau$~Boo B were fixed to those of the astrometric
solution reported by Drummond (2014).}}
\label{fitresults_tauboo}
\begin{tabular}{cc}
\hline
\hline
Parameter & Value\\
\hline
\hline
\multicolumn{2}{c}{$\tau$~Boo b}\\
\hline
\noalign{\smallskip}
Period [days] & $3.3124568\pm 6.9 \times10^{-6}$ \\
T$_{\rm periastron}$ [BJD$_{\rm UTC}$-2450000] & $ 6400.94 \pm 0.30$ \\
K [\ms] & $471.73 \pm 2.97$ \\
$e$ & $0.011\pm 0.006$ \\
$\omega$ [deg] & $113.4\pm  32.2$\\
$\gamma$ [\ms] & $0.0$\\
$m\sin i$ [M$_{\rm Jup}$] & $4.32\pm 0.04 $\\
semi-major axis [AU] & $  0.049 \pm  0.003$ \\
\noalign{\smallskip}
\hline
\multicolumn{2}{c}{$\tau$~Boo B}\\
\hline
\noalign{\smallskip}
Period [years] & $964$ $(fixed)$\\
T$_{\rm periastron}$ [BJD$_{\rm UTC}$-2450000] & $ 12670$ $(fixed)$ \\
K [\ms] & $ 1217.06\pm 222.36$ \\
$e$ & $0.71\pm 0.22$ \\
$\omega$ [deg] & $94.1\pm  64.0$\\
$\gamma$ [\ms] & $-1099.5\pm 273.0$\\
$m\sin i$ [M$_{\odot}$] & $0.4\pm 0.1 $\\
semi-major axis [AU] & $  109 \pm  7$ \\
\noalign{\smallskip}
\hline
\noalign{\smallskip}
offset$_{Lick, setup 13}$ [\ms] &$ 0.0 $\\
offset$_{Lick, setup 6}$ [\ms] &$ -23.17 $\\
offset$_{Lick, setup 8}$ [\ms] &$ 19.03 $\\
offset$_{HARPS-N}$ [\ms] &$ -16294.87 $\\
\noalign{\smallskip}
\hline
\multicolumn{2}{c}{independent long term trends}\\
\hline
\noalign{\smallskip}
RV slope$_{Lick, setup 2}$ [\ms\, y$^{-1}$] &$ 2.7\pm 5.5 $\\
RV slope$_{Lick, setup 13}$ [\ms\, y$^{-1}$] &$ -13.8\pm 5.6 $\\
RV slope$_{Lick, setup 6}$ [\ms\, y$^{-1}$] &$ -8.8\pm 3.4 $\\
RV slope$_{Lick, setup 8}$ [\ms\, y$^{-1}$] &$ -18.6\pm 2.4 $\\
RV slope$_{HARPS-N}$ [\ms\, y$^{-1}$] &$ -36.6\pm 4.0 $\\
\noalign{\smallskip}
\hline
\hline
\end{tabular}
\end{center}
\end{table}

\begin{figure}[!ht]
\begin{center}
\includegraphics[trim = 0.cm 0cm 0.cm 0.cm, clip, width=\linewidth]{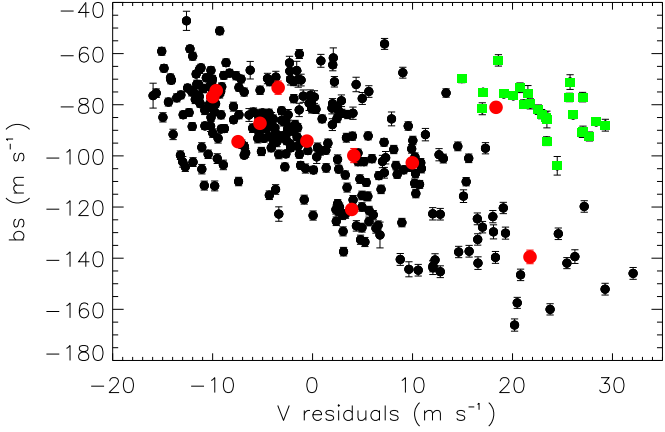}
\caption{\footnotesize{Residuals of the planetary fit for the HARPS-N measurements vs the pipeline-estimated bisector span. The green squares
 refer to exposures taken at JD 2456421, during which the target was observed 
between clouds. Red circles are the measurements for the nightly mean spectra.
Error bars for the bisector span are taken as twice the RV error.
}}
\label{AST01_bs_vs_rv}
\end{center}
\end{figure}

We analyzed the RV residuals shown in the bottom panel of Fig.~\ref{AST01_fit} in frequency, but did not find any trace of additional periodicity.We note in particular that RV residuals of HARPS-N show a clear correlation with
the pipeline-estimated bisector span 
(Fig.~\ref{AST01_bs_vs_rv}), which supports their stellar activity nature (discussed below in Sect. \ref{sec:stellaractivity}).

\subsection{Stellar companion}\label{sect:binary}

Many astrometric measurements have been made to estimate the orbital parameters of the binary stellar system, with the most recent solutions \citep{2011AJ....142..175R,2014AJ....147...65D} claiming that the binary companion has a period $<1000$ years and will approach periastron within the next two decades. 

\begin{figure}[!ht]
\begin{center}
\includegraphics[trim = 0.cm 0cm 0.cm 0.cm, clip, width=\linewidth]{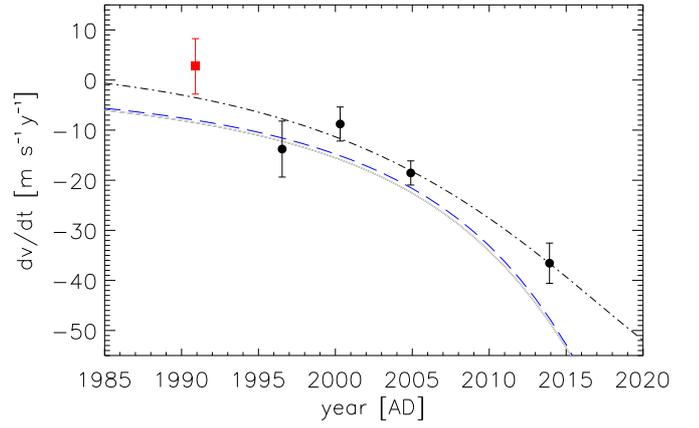}
\caption{\footnotesize{Points represent the values for the long-term RV slopes fitted on Lick and HARPS-N data.
The red square is the value for the first Lick setup with large error bars, which was not taken in consideration when we fit the planetary parameters.
The black dotted-dashed line represents the fitted orbit of the binary companion from the two-objects fit of Table~\ref{fitresults_tauboo}.
The blue dashed line is the long-term slope for the best-fit
orbit for the stellar companion reported by \citet{2014AJ....147...65D}, assuming a mass value of 0.4 $M_{\sun}$. }
}
\label{binary_slope}
\end{center}
\end{figure}

To check whether the RV data match the astrometric best-fit orbit well, we treated each Lick instrumental setup and HARPS-N separately. We fitted a one-planet model, imposing that the orbital solution for $\tau$~Boo b be fixed, and keeping a long-term linear trend free (lower part of Table~\ref{fitresults_tauboo}). In this way, we were able to show the change of the slope in RVs caused by the binary companion $\tau$~Boo B for the first
time.
The information of the astrometric orbital parameters \citep{2014AJ....147...65D} and parallax \citep{2007A&A...474..653V} indicates a mass sum of 1.8$M_{\sun}$ and thus a value of $0.4M_{\sun}$ for $\tau$~Boo B. We used this value to
calculate the astrometric-based orbital RV slope caused in $\tau$~Boo A by the stellar companion (Fig.~\ref{binary_slope}).

Clearly, the data are insufficient for more constraints, but the best-fit astrometric solution agrees well with the present RVs.
The fact that the slope calculated based on HARPS-N data is more than twice those calculated based on Lick data demonstrates that the star is rapidly accelerating as a result of the approaching periastron.
A monitoring of the RV trend during the next years will enable
us to 
obtain more reliable orbital parameters for the stellar binary companion.


\section{Stellar activity\label{sec:stellaractivity}}

In this section we aim to study the SPMI in the $\tau$ Boo system. To 
this purpose, we extensively analyze and compare all the magnetic 
activity indicators available, both computed directly from the spectra and 
extracted from the CCF (see Sect.~\ref{sec:spectralAnalysis}).

\subsection{Analysis of the IRD vs IRD relationships}\label{sec:activity}
\subsubsection{\ion{Ca}{ii} H\&K\ and H$_{\rm \alpha}$}

\begin{figure*}
\centering
\includegraphics[width=.33\linewidth]{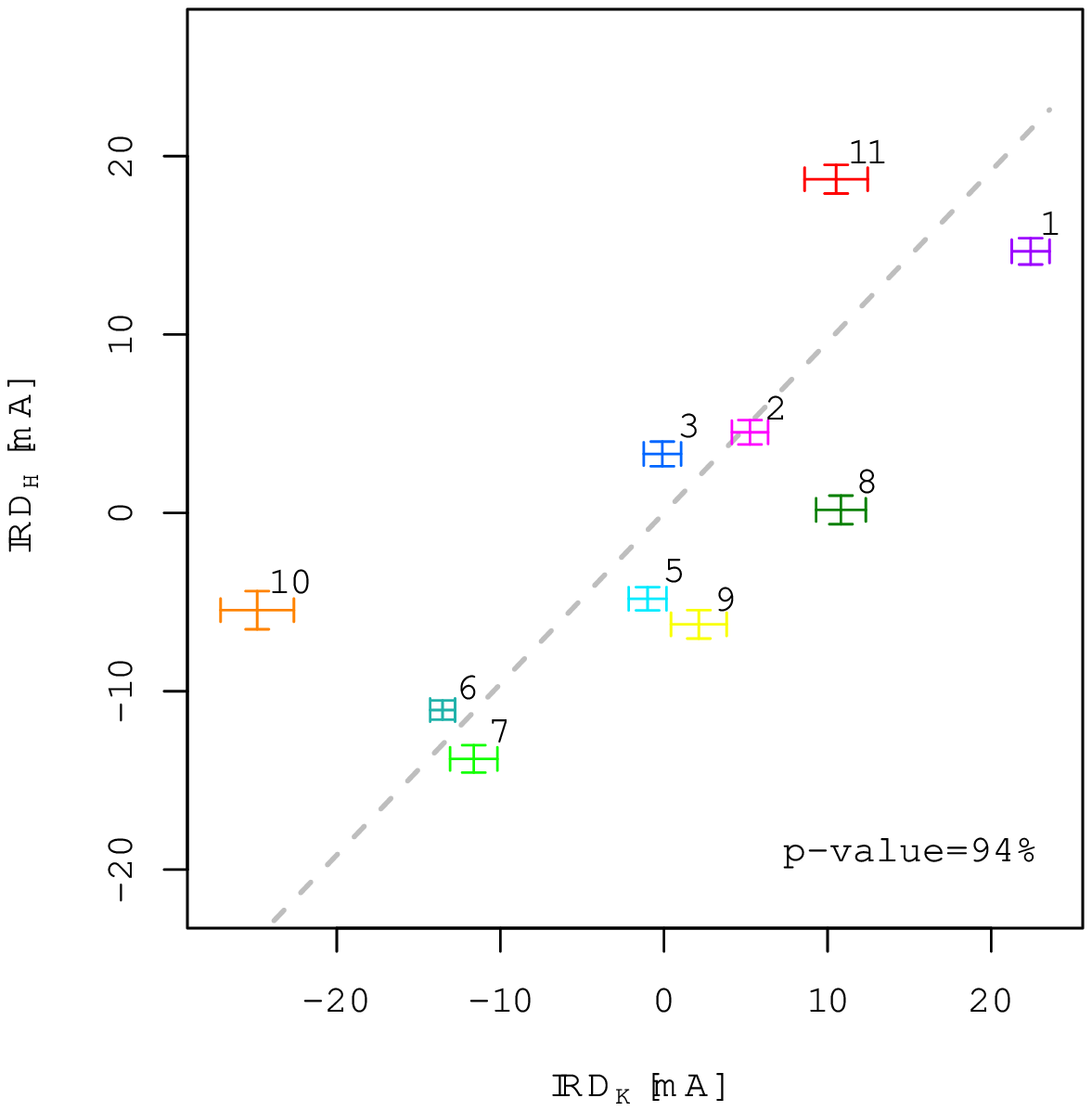}
\includegraphics[width=.33\linewidth]{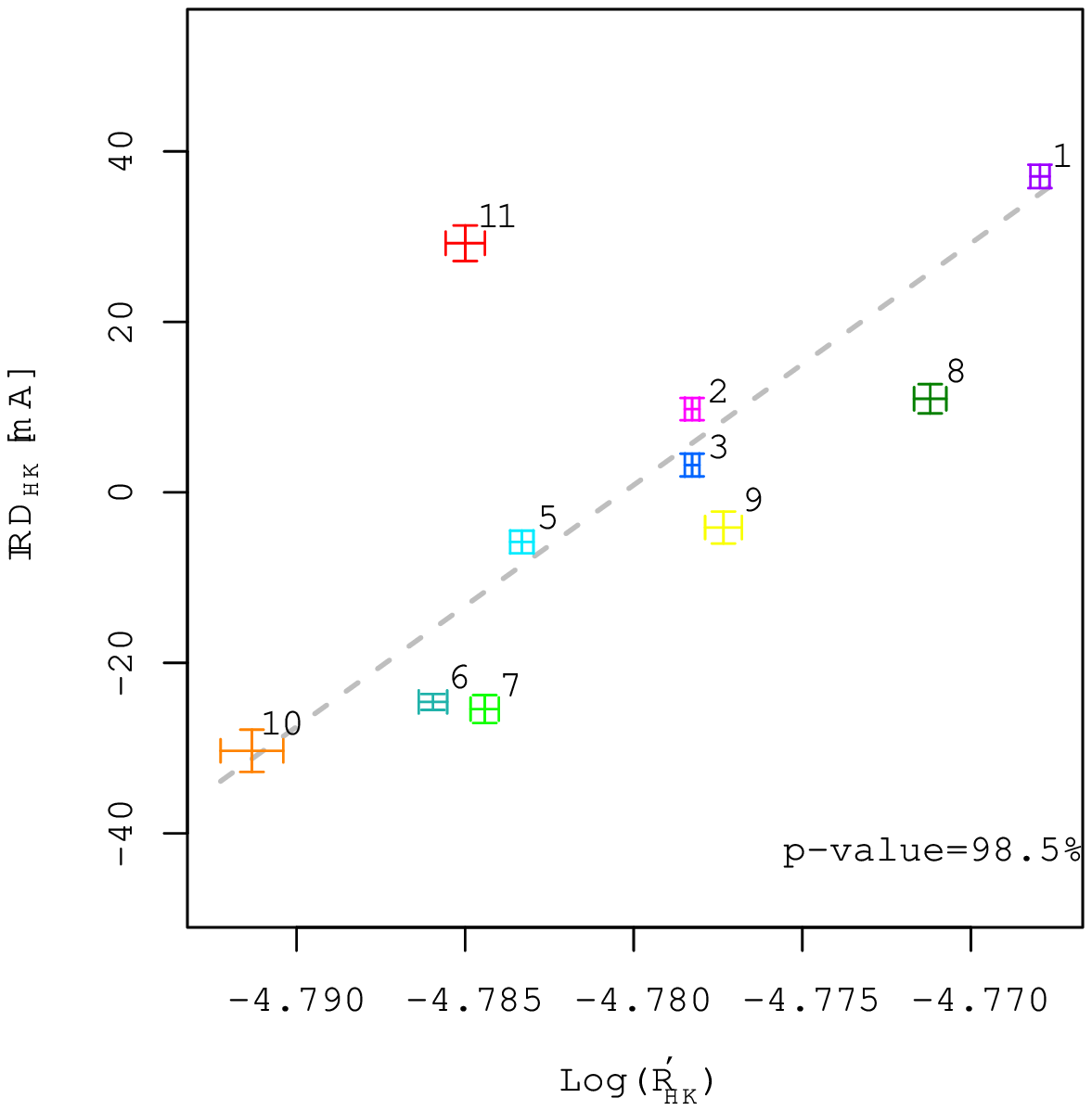}
\includegraphics[width=.33\linewidth]{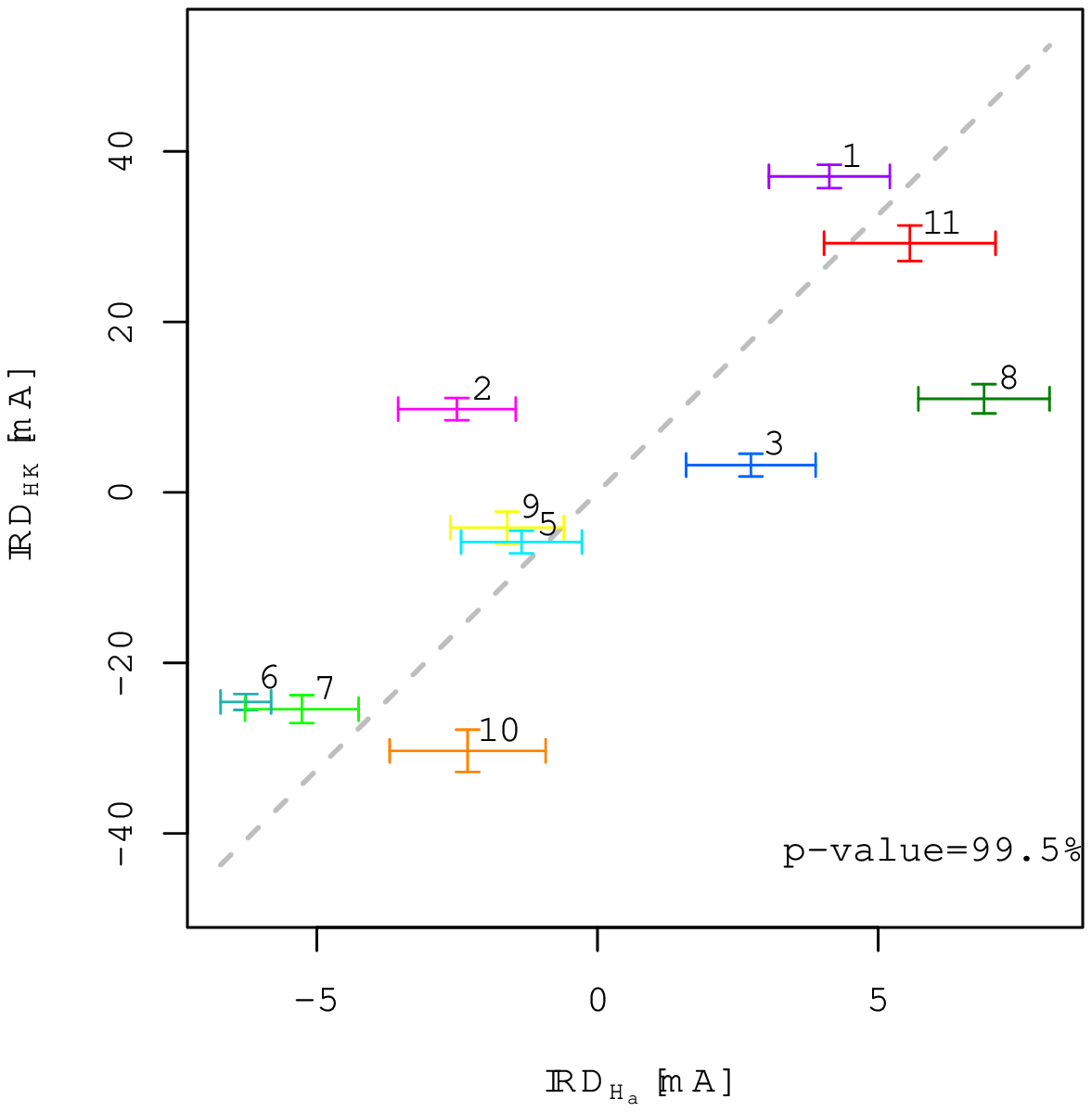}
\caption{\footnotesize{\textit{Left panel: }IRD$_{\rm H}$ vs IRD$_{\rm K}$. \textit{Middle panel: }IRD$_{\rm HK}$ vs Log(R$^\prime_{\rm HK}$). \textit{Right panel: }IRD$_{\rm HK}$ vs IRD$_{\rm H_{\rm \alpha}}$. In all panels, the color code (online version only) is the same as in Fig.~\ref{fig:residuals}, while the number at the bottom is the confidence returned by the correlation test. The numbers above the data points indicate the ordered sequence in the series of observations. Gray dashes show the RMA best-fit.}\label{fig:hk}}
\end{figure*}

Figure~\ref{fig:residuals} shows that the cores of the \ion{Ca}{ii} H\&K\ lines are affected by variability that is stronger than noise in the corresponding spectral regions. In Fig.~\ref{fig:hk} (left panel) we plot the corresponding IRDs of the two lines (IRD$_{\rm H}$ and IRD$_{\rm K}$): we found that the two proxies are strongly correlated with each other with a confidence level of $\sim$94\%; this shows that the detected signal is not an artifact produced by our data reduction, but arises from the magnetic activity of the star. The significance of the correlation was computed with 10000 random permutations of the data, which were also randomly perturbed by the corresponding uncertainties. Since the uncertainties on the variables are similar, the best-fit line was computed by means of a ranged major axis (RMA) regression \citep{Legendre1998}.

Since the two H and K proxies are strongly correlated, we averaged them to obtain the collective IRD$_{\rm HK}$. We found that the latter strongly correlates with the canonical $\log R\arcmin_{\rm HK}$ \citep{1984ApJ...279..763N} computed following \citet{2011arXiv1107.5325L} (Fig.~\ref{fig:hk}, middle panel).

Moreover, as shown in Fig.~\ref{fig:hk} (right panel), IRD$_{\rm HK}$ also strongly correlates with IRD$_{\rm H_{\rm \alpha}}$. A similar result was obtained by \citet{2009MNRAS.398.1383F}, who spectroscopically monitored the activity level of $\tau$ Boo in 2008. This is consistent with different studies by other authors \citep{2009A&A...501.1103M,2010A&A...520A..79M,Martinez2011,2013A&A...558A.141S,2014A&A...566A..66G}, who found strong pairwise correlations between the H{\&}K and the H$_{\rm \alpha}$ lines despite the different formation heights (lower and upper chromosphere) and adopting different approaches (e.g., single-epoch comparison of many stars, multi-epoch variability study of single objects, or emission line fluxes vs indices as relative measures).

\subsubsection{\ion{Na}{i} D$_{1,2}$ doublet and \ion{He}{i} D$_3$ triplet}

\begin{figure*}
\centering
\includegraphics[width=.33\linewidth]{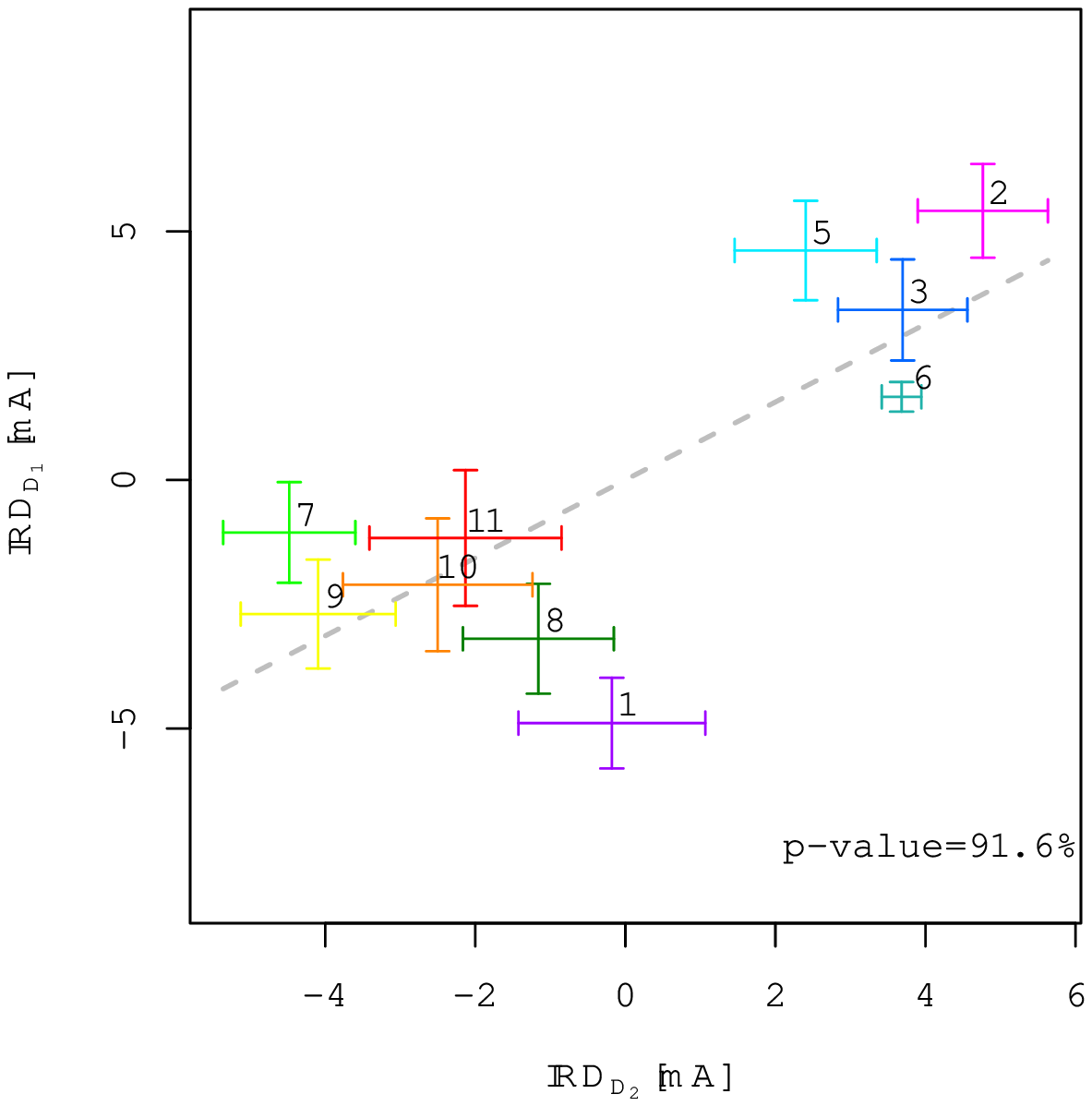}
\includegraphics[width=.33\linewidth]{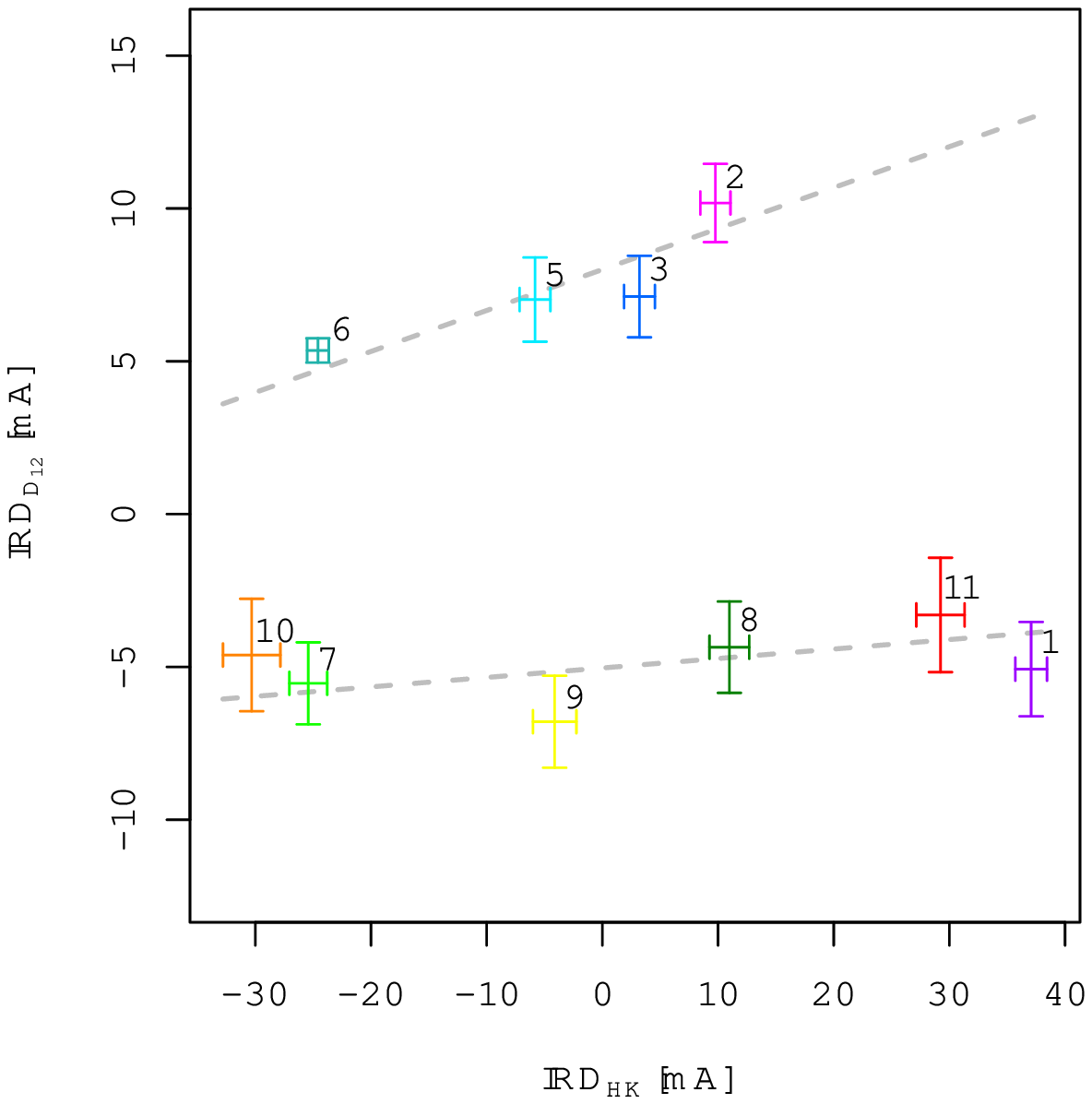}
\includegraphics[width=.33\linewidth]{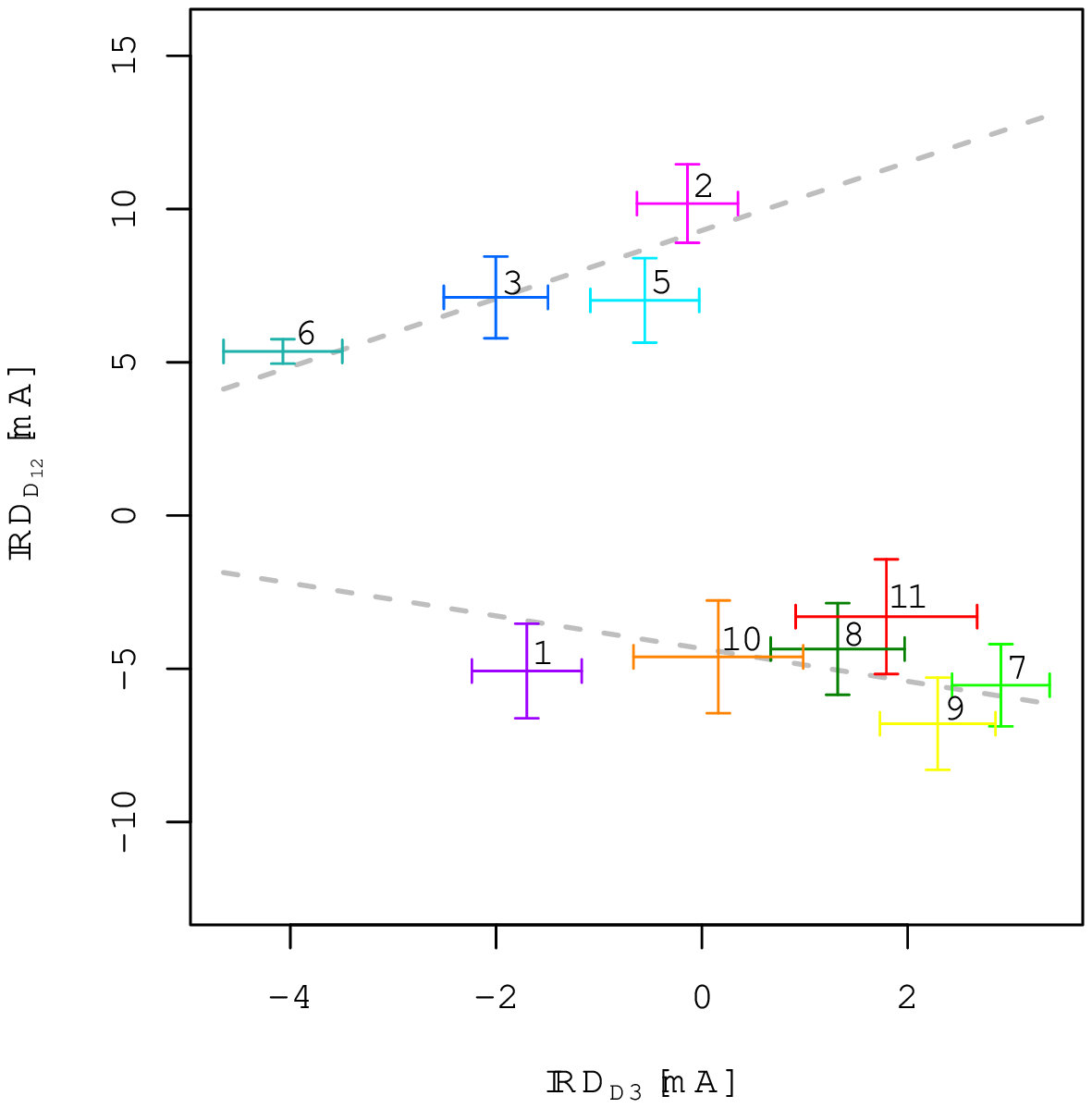}
\caption{\footnotesize{\textit{Left panel: }IRD$_{\rm D_1}$ vs IRD$_{\rm D_2}$. \textit{Middle panel: }IRD$_{\rm D12}$ vs IRD$_{\rm HK}$. \textit{Right panel: }IRD$_{\rm D12}$ vs IRD$_{\rm D3}$. In all panels, the color code (online version only) is the same as in Fig.~\ref{fig:residuals} and represents different observation dates. The numbers above the data points indicate the ordered sequence in the series of observations.  In the left panel, the number at the bottom is the confidence returned by the correlation test. In the middle and right panels, the two dashed gray lines are drawn by hand to highlight the bifurcation discussed in the text.}\label{fig:nai}}
\end{figure*}

We found that the IRDs of the \ion{Na}{i} D$_{\rm 1}$ and \ion{Na}{i} D$_{\rm 2}$ lines correlate linearly, but with a confidence level not better than $\sim$92\% (Fig.~\ref{fig:nai}, left panel). We carefully checked the spectra and did not find any hint of a poor telluric correction or uncorrected terrestrial emission lines above the noise level. As described above for the H{\&}K lines, we averaged the two proxies to obtain the IRD$_{\rm D12}$ indicator.

We did not find any correlation from plotting IRD$_{\rm D12}$ against IRD$_{\rm HK}$ (Fig. ~\ref{fig:nai}, middle panel). This is consistent with the results of \citet{2007MNRAS.378.1007D}, who only found a good correlation between the Na and Ca doublets for stars with Balmer lines in emission (not the case of $\tau$ Boo). In any case, it is interesting to note that there is a bifurcation in the sample of measurements (Fig.~\ref{fig:nai}, middle panel). 

In a similar way, the IRD$_{\rm D12}$ vs IRD$_{\rm D3}$ plot seems to cluster into two subgroups (Fig.~\ref{fig:nai}, right panel). The origin of the two trends is still unclear. \citet{1981ApJ...251..768L}, analyzing a few quiescent prominences on the Sun, obtained a similar bifurcation by comparing the intensities of the \ion{He}{i} D$_3$ triplet with the \ion{Na}{i} D$_1$ and D$_2$ lines. He conjectured, with no clear demonstration, that since the He triplet is back-heated by coronal UV radiation, then the two branches reflect the association of prominences with nearby coronal activity of different levels.
The author remarked that the origin of the two branches is related to the \ion{He}{i} D$_3$ triplet, as it shows the same bifurcation when compared to the intensity of the \ion{Ca}{ii} $\lambda$8498 line, while the latter well correlates with the \ion{Na}{i} D$_{12}$ lines. In contrast, we found the bifurcation when IRD$_{\rm D12}$ was compared with both IRD$_{\rm D3}$ and IRD$_{\rm HK}$, while there is no bifurcation in the IRD$_{\rm D3}$ vs\ IRD$_{\rm HK}$ plot. Thus our results indicate that the formation of the \ion{Na}{i} doublet should be investigated to explain this twofold behavior.

Another possibility is the presence of prominence-like structures around the stars, formed with matter evaporating from the planet and supported by the magnetic field of the star \citep{2014A&A...572L...6L}. Since exoplanets are expected to be richer in metal than stars \citep[e.g., ][]{2006ApJ...642..495F}, this may explain why \ion{Na}{i} shows the two branches instead of the \ion{He}{i} D$_{\rm 3}$ triplet.

Focusing on the \ion{He}{i} D$_{\rm 3}$ triplet, we found that the scatter of the IRD$_{\rm D3}$ measurements is slightly larger than the uncertainties, indicating that weak variability in this proxy has occurred, if any. Moreover, we did not find the correlation with the \ion{Ca}{ii} H\&K\ lines that has previously been found by \citet{1993A&A...273..482G} in their sample of F-type main-sequence stars. This is probably because their statistical sample spans a wide range of $\log R\arcmin_{\rm HK}$, while $\tau$ Boo's chromospheric variability remained too limited during our observational campaign to detect such a correlation.

\subsubsection{Parameters of the CCF}

The very high S/N of the mean spectra allowed us to study the CCF variability with a high degree of accuracy. In particular, since the CCF is obtained from a large set of photospheric lines, the CCF variability is indicative of the source and level of photospheric activity. It is thus interesting to study the CCF profile to understand the phenomenology of magnetism on the surface of $\tau$ Boo and how it is related to chromospheric activity.

As suggested by \citet{Nardetto2006}, one way to analyze the CCF profile is to retrieve its width and contrast to the continuum by means of a Gaussian best-fit. They also introduced the possibility to fit asymmetric profiles by tweaking the Gaussian model. For $\tau$ Boo, both methods led to poor fits of the observed CCF profiles, probably because the differential rotation combined with the rotation rate of the star broadens the line profile in a non-Gaussian way.

We therefore approached the CCF profile analysis by averaging the CCFs returned by the DRS from the series of ten nightly averaged spectra, obtaining a master CCF profile with a S/N higher than each nightly averaged CCF's S/N. Then, we fitted each nightly
averaged CCF with the master CCF, allowing for changes in FWHM and contrast (CCFc).

We also computed the BIS and the v$_{asy}$ parameters defined by \citet{Figueira2013}. The former measures the velocity difference between the midpoints at the top and bottom of the CCF \citep{Gray2008}, while the latter gives an estimate of the asymmetry of the CCF profile. By definition, the computation of both parameters is model independent.

Since \citet{2001A&A...379..279Q}, variations in BIS are the paradigm for the signature of activity-induced RV variations. These authors found a strong anticorrelation between the measured RVs and the corresponding BIS of the CCF. Since then, the anticorrelation between BIS and residual radial velocities (RV$_{res}$, the RVs minus the planetary best-fit orbit) is considered a clear signature of short-term (i.e.,\ rotationally induced) photospheric variability. In our case, the BIS shows a high degree of anticorrelation with the residuals RV$_{res}$ (Fig.~\ref{AST01_bs_vs_rv}). This suggests that RV$_{res}$ are due to stellar activity.

\begin{figure*}
\centering
\includegraphics[width=.33\linewidth]{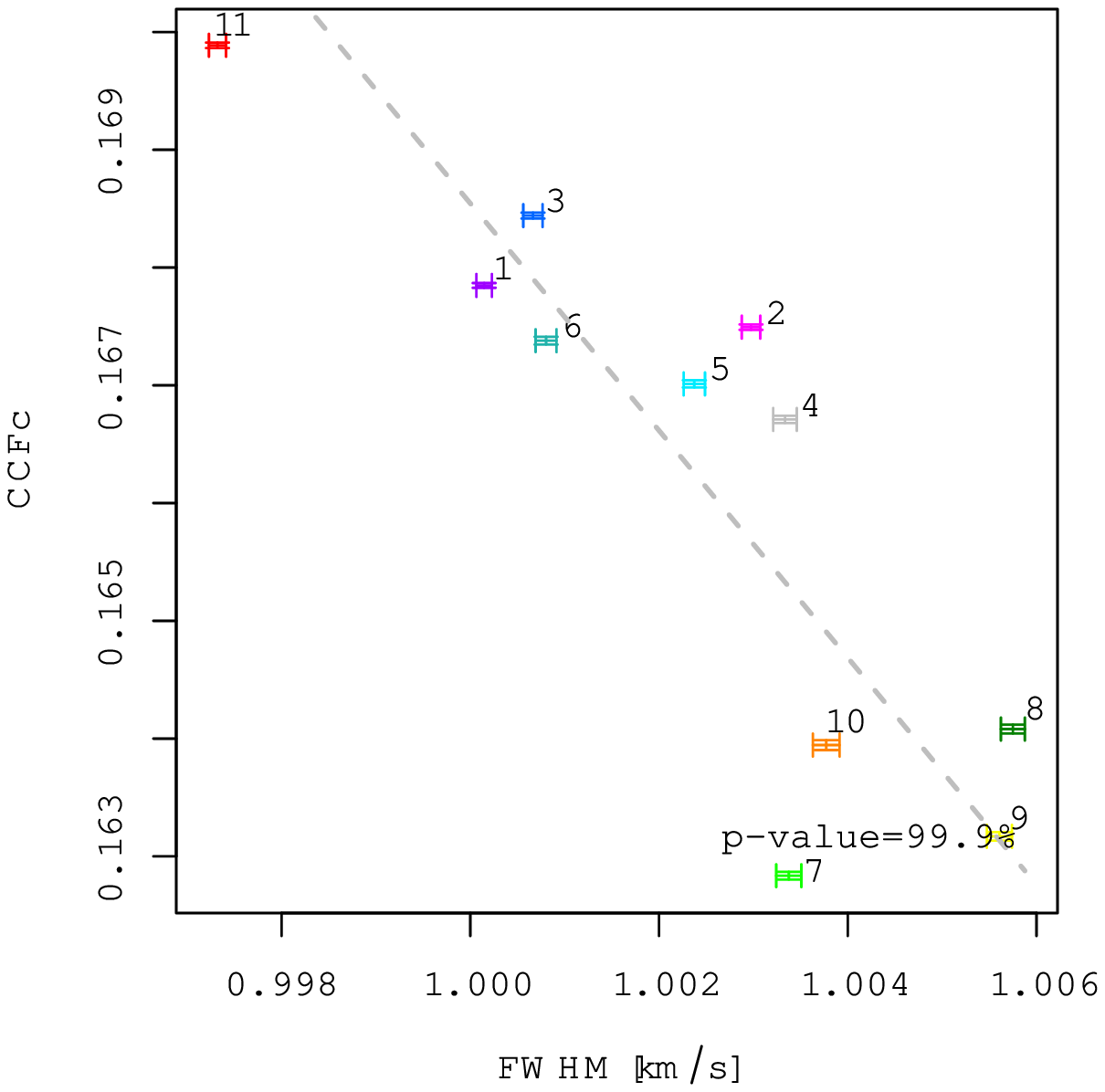}
\includegraphics[width=.33\linewidth]{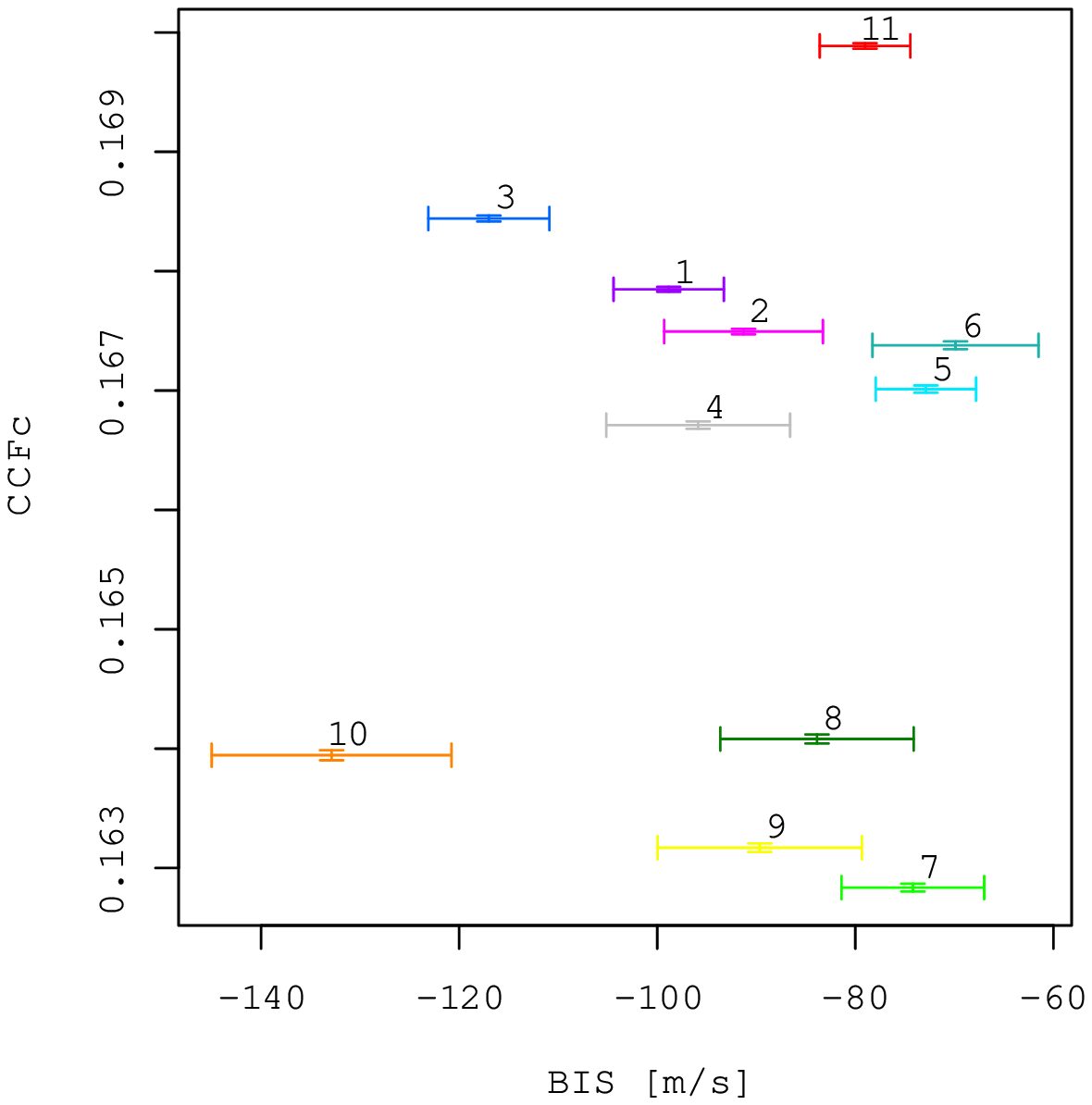}
\includegraphics[width=.33\linewidth]{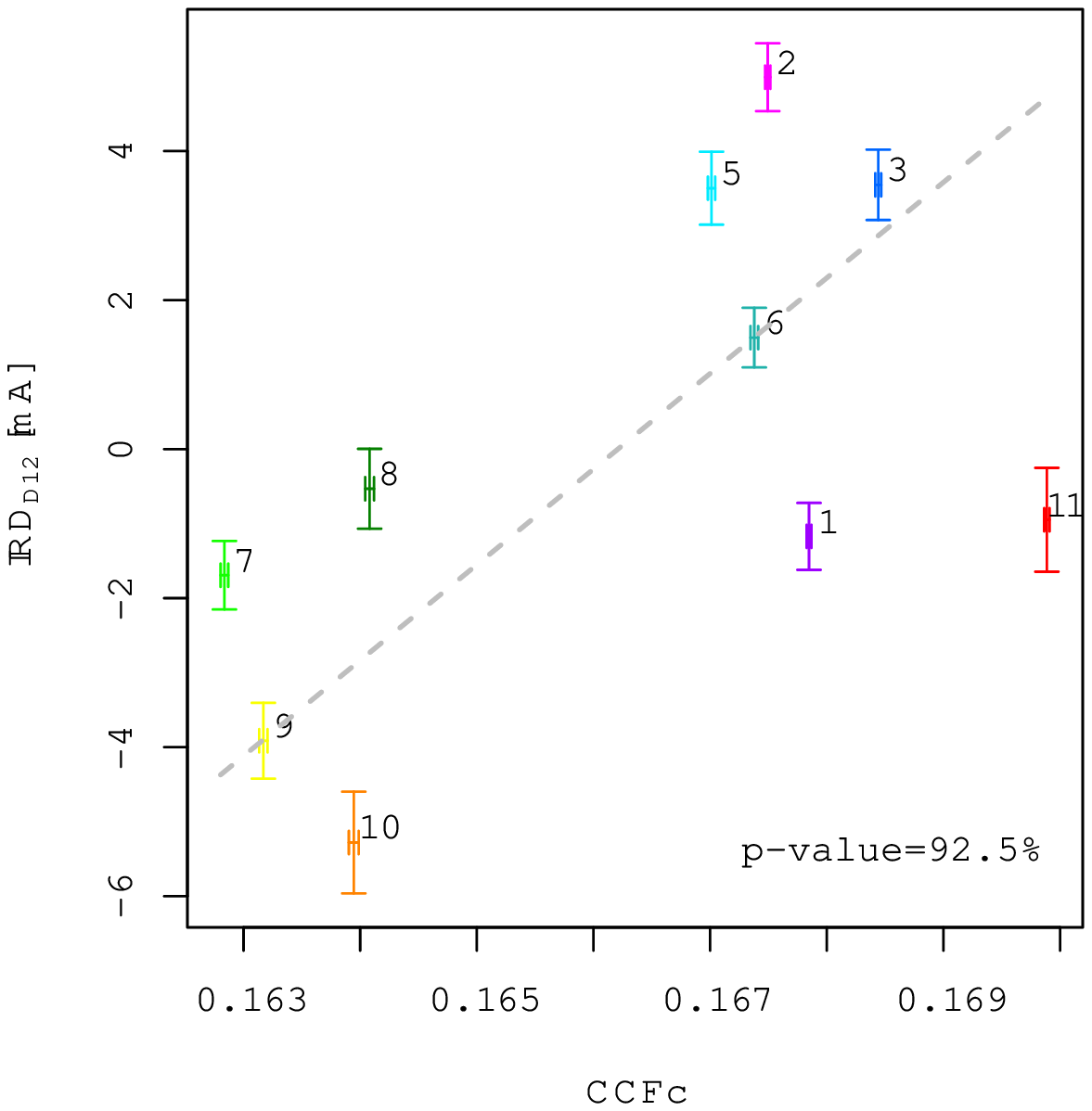}
\caption{\footnotesize{\textit{Left panel: }CCFc vs FWHM.  \textit{Middle panel: }CCFc vs BIS. \textit{Right panel: }IRD$_{\rm D12}$ vs CCFc. Symbols and colors (online version only) are the same as in Fig.~\ref{fig:residuals}.}\label{fig:ccfcvasy}}
\end{figure*}

Some other parameters show a strong anticorrelation: CCFc and FWHM (Fig.~\ref{fig:ccfcvasy}, left panel). This is consistent with the fact that while the FWHM measures the width of the CCF at half maximum, CCFc counteracts this to preserve the area of the CCF. The two indices are thus sensitive to the broadening of the wings of the CCF. We infer, accordingly, that correlations between these indices suggest a broadening variability of the CCF.

Our results therefore indicate that the BIS follows the deformations of the CCF better than CCFc and FWHM which, in turn, are sensitive to the overall width of the CCF. The BIS and CCFc can then be considered the best representatives of two different families of indicators. Given the operational definitions of the BIS and CCFc, we can regard the former as an indicator of the mean distortion of the line profile and the latter as an indicator of the mean strength of the photospheric lines.

The weak variability of BIS with respect to measurement uncertainties compared to CCFc (Fig.~\ref{fig:ccfcvasy}, middle panel) may give some clues on the geometry of active regions on the stellar surface. The stability of the BIS and the variability of CCFc may indicate that the CCF changes its contrast with respect to the continuum while preserving its overall asymmetry. This is consistent with a scenario with an active region around the pole of the star, which is not Doppler-shifted by stellar rotation and rapidly evolves in terms of contrast to the quiet photosphere (either in temperature difference and/or coverage factor).

We did not find any significant correlation between v$_{asy}$ and the other parameters of the CCF.

When we compared the CCF parameters with the chromospheric indicators discussed in this section, we found the strongest correlation between IRD$_{\rm D12}$ and CCFc (p=92.5\%, right panel in Fig.~\ref{fig:ccfcvasy}). The same applies for IRD$_{\rm HK}$ and IRD$_{\rm H_{\rm \alpha}}$ with a lower confidence level (p$\lesssim$90\%). We infer that the variability of the \ion{Na}{I} D$_{12}$ doublet arises from changes in the line formation physics at the photospheric level, rather than from genuine chromospheric emission. The doublet is formed in the lower chromosphere \citep{Tripicchio97} and is thus more closely related to the magnetic activity of the lowest chromosphere and photosphere. Moreover, as stated above, the \ion{Na}{I} D$_{12}$ doublet is a good proxy for chromospheric activity only for mid-to-late type stars with emission Balmer lines, which is not the case of $\tau$ Boo.

\subsection{Time series analysis}\label{sec:time}

\begin{figure*}[!ht]
\centering
\includegraphics[width=.33\linewidth]{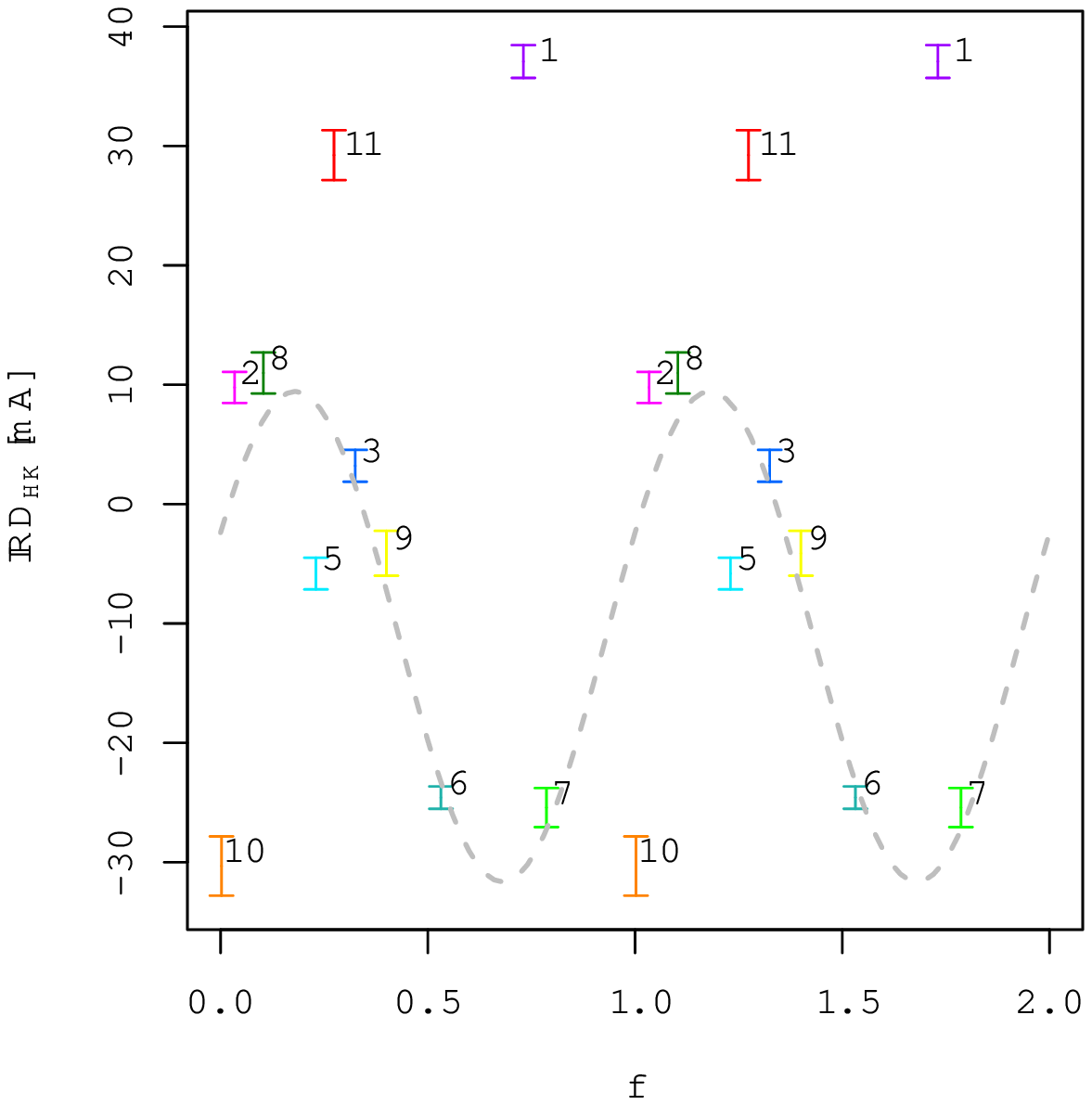}
\includegraphics[width=.33\linewidth]{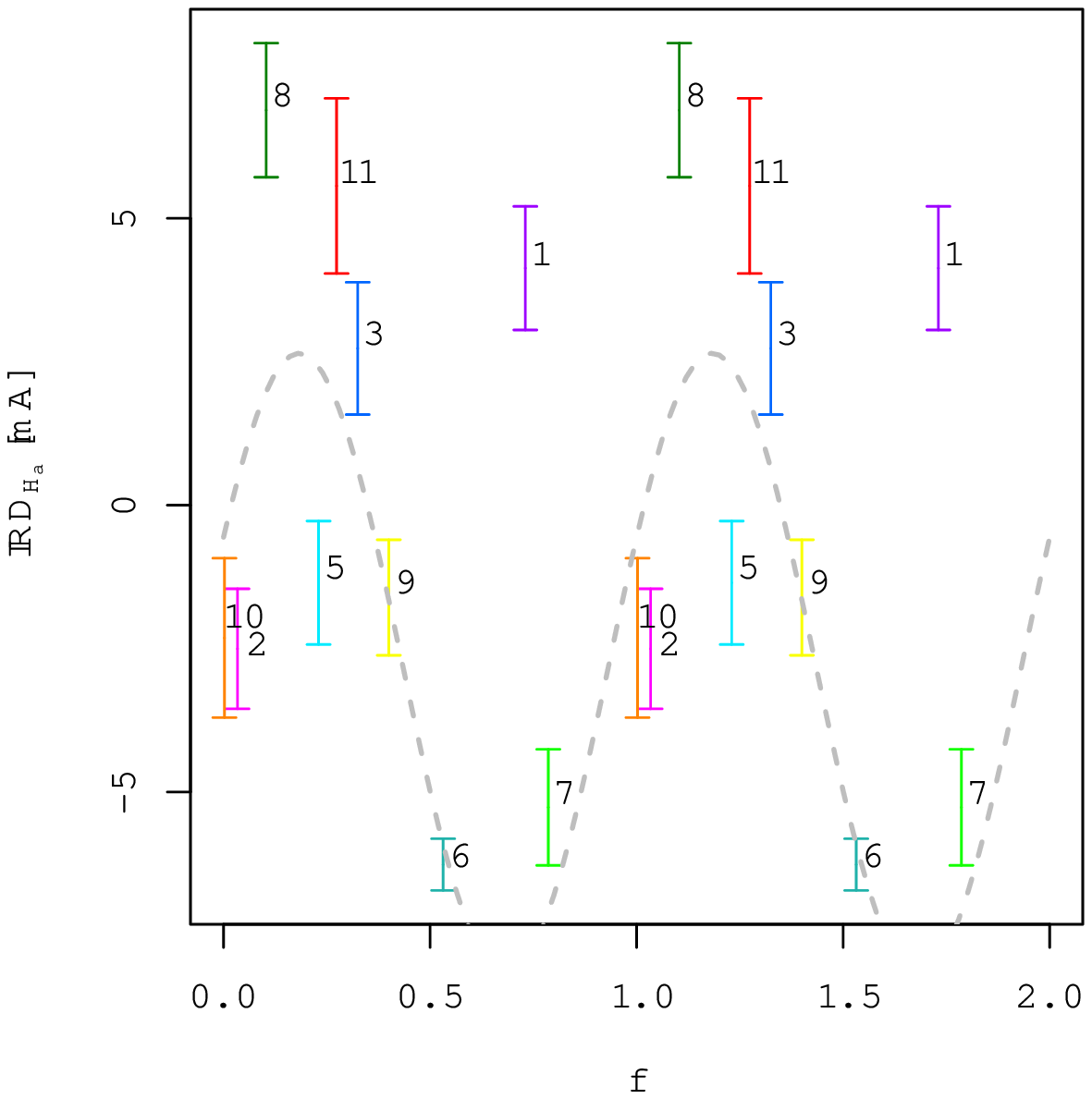}
\includegraphics[width=.33\linewidth]{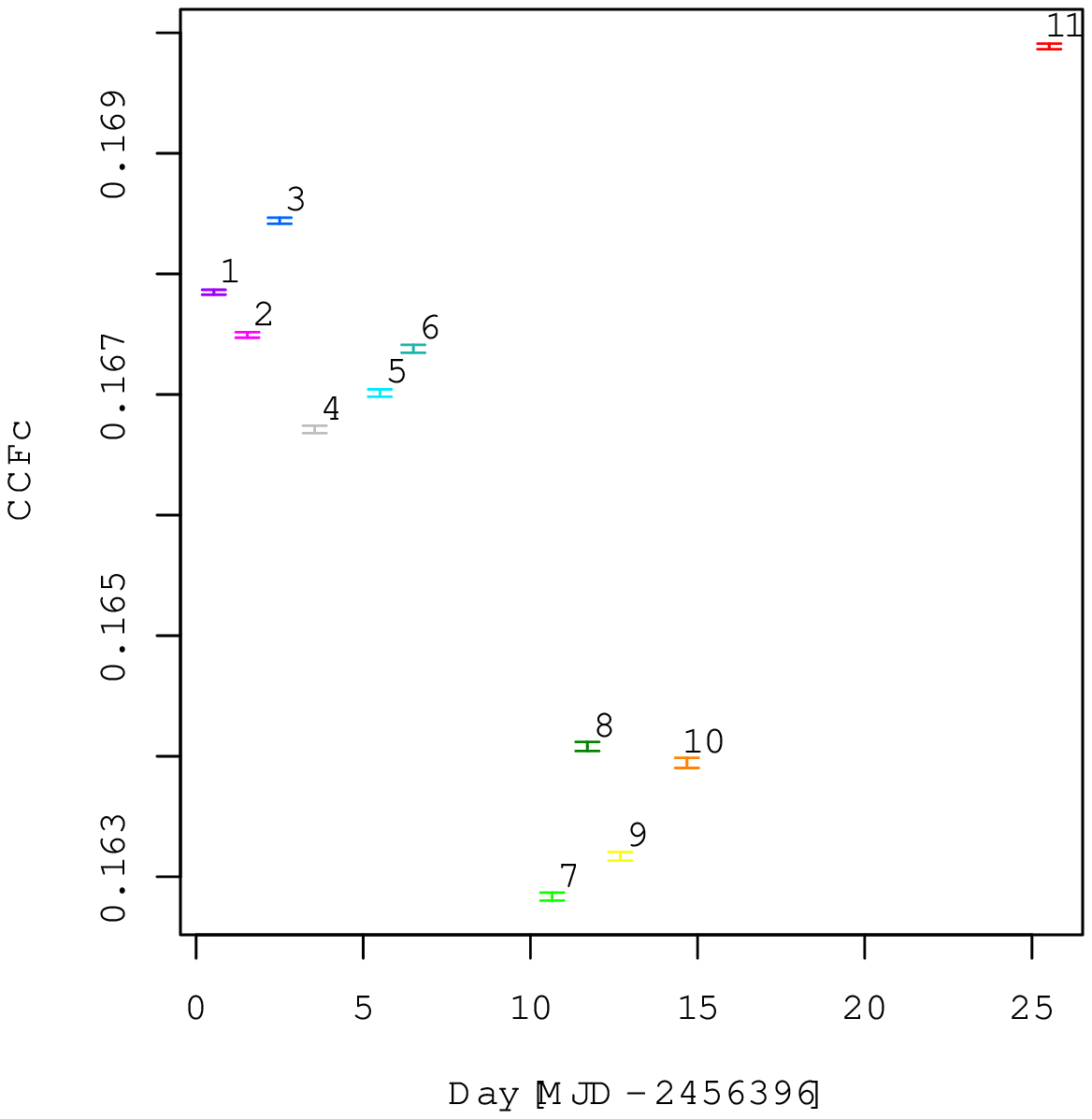}
\caption{\footnotesize{\textit{Left panel: }Phase-folded diagram of IRD$_{\rm HK}$. \textit{Middle panel: }IRD$_{\rm H_{\rm \alpha}}$. \textit{Right panel: }Time series of the CCFc measurements. The planetary inferior conjunction is at $\phi$=0. Colors (online version only) and numbers are the same as in the previous figures. In the left and middle panel, the gray dashed line is the least-squares best fit discussed in the text. We also report in the text the relative likelihood of the sinusoidal fit with respect to the flat model returned by the AICc.}\label{fig:phaseFolded}}
\end{figure*}

Using the ephemeris of $\tau$ Boo b reported in Table~\ref{fitresults_tauboo}, we searched for the eventual phasing of each of the indicators discussed so far with the orbital motion of the planet. The most convincing cases are the phase-folding of IRD$_{\rm HK}$ and IRD$_{\rm H_{\rm \alpha}}$ shown in Fig.~\ref{fig:phaseFolded} (left and middle panel), where we set the phase $\phi=0$ at the planetary inferior conjunction. In these diagrams we overplot the weighted least-squares best fit of the form IRD=A$\cdot\sin2\pi\phi$+B$\cdot\cos2\pi\phi$+C, where $\phi$ is the orbital phase and A, B, C are the parameters to be fitted. We excluded JD 2456396 (i.e.,\ night 1) from the fit because it clearly is an outlier. The coefficient of determination $R^2$ of both fits is $\sim$50\%, which means that\ the model explains $\sim$50\% of the total variance of the sample. Moreover, the likelihood ratio test suggests that the sinusoidal fit is to be preferred with respect to the flat model with a relative likelihood close to 100\%.

For all the other indicators, both chromospheric and CCF-related, we obtained a lower $R^2$ coefficient, that is,\ we found weaker or no evidence of such a phasing. Conversely, we found that CCFc, and consequently IRD$_{\rm D12}$ , show a clear trend with time (right panel in Fig.~\ref{fig:phaseFolded}). If we interpret that CCFc anticorrelates with magnetic activity \citep[either spot-like or plage-like, see][]{2014ApJ...796..132D}, then the decrease of CCFc between JD 2456402 and 2456407 suggests that $\tau$ Boo increased its activity level, and conversely it turned over toward a quieter configuration between JD 2456411 and 2456422.

\subsection{Discussion}\label{sec:activityDiscussion}

In Sect.~\ref{sec:activity} we argued that the CCFc shows variations on time scales longer than the stellar rotation. Polar magnetic regions on $\tau$ Boo have already been detected in several epochs by \citet{Fares2013}, who remarked that the polarity of the magnetic field switches $\text{about every two}$ years. Moreover, inside each activity cycle they found signatures of a rapid evolution of $\tau$ Boo's magnetic field. This supports the hypothesis that the variations of CCFc are driven by a large-scale evolution of magnetic activity, even along time spans as short as $\sim$20~days (Fig.~\ref{fig:phaseFolded}, right panel).

Previous photometric monitorings of the star have never detected photometric variability above the mmag level \citep{1997ApJ...474L.119B,2008A&A...482..691W}. Accordingly, we assume that the typical peak-to-peak variability is on the order of 1~mmag (upper limit). With this assumption, and assuming that the active region is spot-dominated,  we obtain that the active region covers $\sim$0.1\% of the visible hemisphere using
the SOAP2.0 code \citep{2014ApJ...796..132D}. The corresponding peak-to-valley variations of RV, BIS, and FWHM returned by the simulations are $\lesssim$25~\ms, while those of our measurements are actually larger ($\sim50$~\ms, $\sim$70~\ms, and $\sim$150~\ms\,respectively).

Conversely, if we assume that the active region consists of a bright plage, then the coverage factor is $\simeq$2.5\%. With this coverage, the simulated activity-induced signal in RV, BIS, and FWHM is on the order of 100~\ms, comparable with those of our measurements. We thus conclude that the high-latitude active region probably is plage-dominated.

In Sect.~\ref{sec:time} we briefly discussed the time variability of the most representative indicators, finding that the genuine chromospheric indicators, such as IRD$_{\rm HK}$ and IRD$_{\rm H_{\rm \alpha}}$, seem to be phased with the orbital motion of the planet. This has previously been found by \citet{2008A&A...482..691W}, who detected a plage on $\tau$ Boo at planetary phase $\phi\simeq 0.8$ from \ion{Ca}{ii} H{\&}K spectra in 2001 to 2003 and claimed a photospheric spot using MOST photometric data taken in 2004 and 2005. The authors stated that the persistence of the active region at the same longitude is a strong indication that it is caused by a magnetic link between the planet and the star. On the other hand, \citet{Mathur2014} have recently reported a few cases of bona fide single F-type stars observed with {\it Kepler} with active longitudes persisting on the stellar surface for many stellar rotations.

Since $\tau$ Boo's rotational period equals the planetary period \citep[with some degree of differential rotation, ][]{2009MNRAS.398.1383F}, it is still unclear whether the phasing of chromospheric activity with the planet is due to SPMI or to 
an active region that simply corotates. 
In any case, if the SPMI scenario is confirmed, we remark that in our observing season the plage was located at $\phi\sim$0.1-0.2 (Fig.~\ref{fig:phaseFolded}), indicating that the magnetic connection between planet and star, if present, has moved between years 2005 and 2013. This may be due to a poloidal reversal of $\tau$ Boo's magnetic fields between these epochs. 
However, we remark that at phase 0.8 we found the largest scatter in the activity diagnostics:
we have only two observations, and no statistical significance of this hypothesis can be assessed.


\section{Asteroseismology\label{sec:seismology}}

We have shown in the previous sections how the mean spectra can be used to derive 
very precise radial velocity measurements to study the planetary and binary
orbits and very accurate indicators to monitor the stellar activity and 
star-planet interaction. We now describe the results obtained from the
analysis of the high-cadence, short-exposure HARPS-N spectra. 

\begin{figure*}[!ht]
\begin{center}
\includegraphics[trim = 0cm 0cm 0cm 0.1cm, clip, width=16cm]{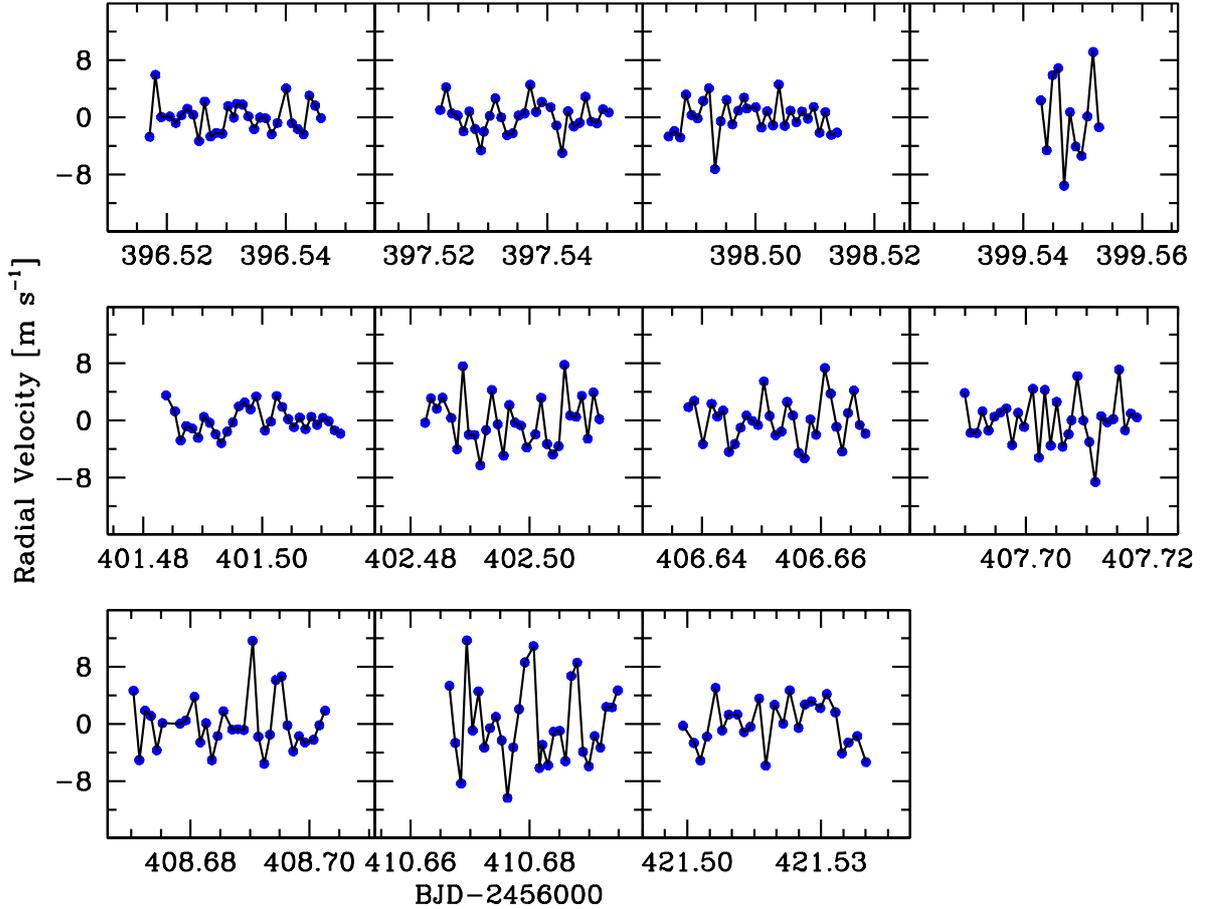}
\caption{\footnotesize{Asteroseismology of $\tau$~Boo:  a linear fit was subtracted from
the original HARPS-N data after the detrending from the planetary and binary orbits.}}
\label{tauboo_ex_sismo}
\end{center}
\end{figure*}

We started our asteroseismic analysis with correcting the Doppler shift that is due to the orbital motion of the planet (Sect.~\ref{OrbitalFitSect}).
Then, to remove the residual low-frequency term left in the data,
several approaches were tried, and we report the most powerful approach here: 
the subtraction of a linear trend from the data of each night. This procedure ensured that no spurious low-frequency term (due to instrumental problems, planetary and binary orbits,
or RV variations induced by activity phenomena with timescales longer than the duration of one
observing sequence) affects the radial velocity measurements that were used for the asteroseismic analysis.

This procedure returned time series centered on RV=0.0~\ms\, on each night
(Fig.~\ref{tauboo_ex_sismo}). The resulting RV curves show sporadic peak-to-peak variations  of up to 
 15~\ms\, (JD~2456399 and 2456410), but $\tau$~Boo typically
is much quieter, at lower than 10~\ms. 
This amplitude is larger than that of the solar twin 18~Sco \citep{18sco}, but comparable 
with that of the subgiant $\beta$~Hyi  \citep{betahyi}.

\begin{figure}[!ht]
\begin{center}
\includegraphics[width=\linewidth]{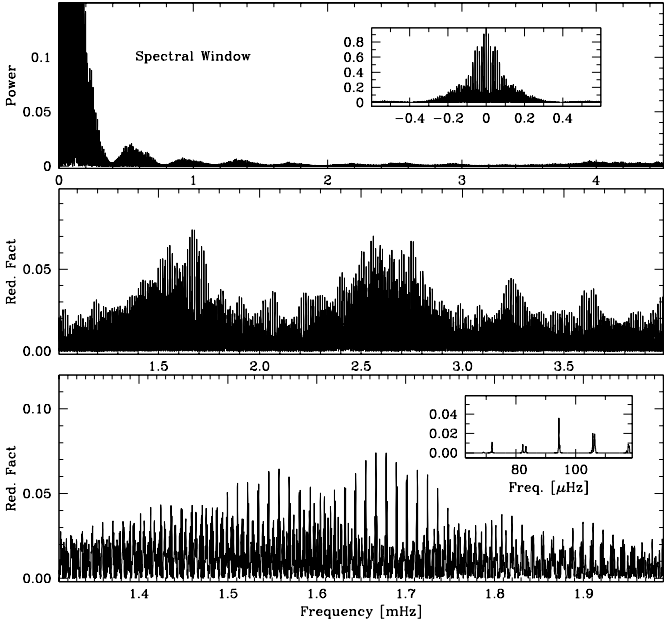}
\caption{\footnotesize{Asteroseismology of $\tau$~Boo.
\emph{Top panel}: spectral window of the HARPS-N data in the 0.0-4.5~mHz range and zoom in the 0.0-0.5~mHz range (inset).
\emph{Middle panel}: power spectrum in the 1.0-4.0~mHz range.
\emph{Bottom panel}: power spectrum in 1.3-2.0~mHz range and power spectrum of this spectrum (inset). }}
\label{tauboo_sismo}
\end{center}
\end{figure}

The time series 
were then analyzed in frequency
to detect short-scale periodicities. We used the iterative sine-wave least-squares method
\citep{vani} and checked the results with the generalized Lomb-Scargle periodogram \citep{scargle}. 
The power spectra  are very similar and strongly 
affected by the sampling of the observations. The spectral window shows strong aliases at multiples of
1~\cd, i.e., 11.57~\mhz\, (Fig.~\ref{tauboo_sismo}, inset in the top panel). 

As expected because of the limited time coverage, the frequency analysis of the RV values cannot supply
a unique determination of the asteroseismic content. 
The power spectrum is very noisy: a first relevant pattern occurs between 1.5 and 1.8~mHz, followed 
by another between 2.4 and 2.9~mHz (middle panel). 
Both are affected by the spectral window effects: each pulsational mode of $\tau$~Boo originates a structure
similar to the spectral window, destroying the expected comb structure of the excited modes.
The former pattern stands out a little more clearly over the noise, and it shows the highest peaks
around 1.68~mHz (bottom panel). The corresponding period of  9.9~min can be glimpsed
in Fig.~\ref{tauboo_ex_sismo}: it is traced by six consecutive measurements (60 sec exposure
time, 25-35 sec overhead), but of course often modified by the interference with the other modes.
The amplitudes of the peaks are very small, around 1.1~\ms. This value takes into account the intrinsic
incoherence (amplitude damping, mode lifetimes) of the solar-like oscillations. 
We also calculated the level of the noise in the RV time series and  
obtained 0.35~\ms. The corresponding S/N=3.1 leaves some uncertainties on the significance of the 1.5-1.8~mHz pattern 
since the threshold S/N=4.0 \citep[e.g., ][]{betahyi} was not reached.
All these observational uncertainties are due to the limited time coverage:
twice as many measurements would have allowed us to reach a
threshold of  S/N=4.0.

\subsection{Asteroseismic results: observation vs theory}
The availability of a small set of high-precision radial velocity measurements
allowed us to obtain an estimate of the asteroseismic parameters of $\tau$~Boo
making a very limited observational investment. We verify these results
in  a theoretical context.

We computed the expected place of an excess of power in a star like $\tau$~Boo.
To do this, we used the scaling relations between stellar parameters ($M$, T$_{\rm eff}$, and $L$; Table~1) and 
the frequency of maximum power of the oscillations $\nu_{\rm max}$ \citep[e.g.,][]{stello}.
We obtained 1.98$\pm$0.46~mHz, compatible at $\pm1\sigma$ level with the observed value.
We performed another check on
the detection of $\nu_{\rm max}$ by computing the
power spectrum of the power spectrum to identify regularities in the detected frequencies. The most
relevant feature was the 11.57~\mhz\, spacing that is caused
by the aliasing effect. After this, a peak
at 94.4~\mhz\, has appeared (inset in the bottom panel of Fig.~\ref{tauboo_sismo}).
 If the structure centered at 1.68~mHz 
is due to solar-like oscillations, 
then the 94.4~\mhz\, spacing should be the large separation. The
pair ($\nu_{\rm max}$, $\Delta\nu$)=(1680, 94)~\mhz\, matches the observed relation 
\citep[see Fig.~2 in][]{stello}. Although not yet decisive, these results give us more confidence
in the asteroseismic approach to study the stars hosting exoplanets by means of radial

velocity measurements performed mainly for other purposes.


\section{Evolutionary stage\label{sec:evolution}}

It has been well demonstrated  \citep[e.g.,][]{cunha} that an
accurate determination of mass 
and radius combined with the spectroscopic measurements of atmospheric parameters allows us to clearly constrain the stellar structure of observed stars.

To assess the information available from HARPS-N observations,
we therefore calculated a grid of theoretical structure models of $\tau$~Boo evolved from a chemically uniform model on the zero-age main sequence,  using the ASTEC evolution code 
\citep{chris2008a} and varying the mass and the composition to match the available atmospheric parameters (see Table~\ref{stellar_param}).

The evolutionary models have been produced by employing current physical information 
following the procedure described in \citet{dimauro2011}.
The input physics for the evolution calculations included the OPAL 2005 equation of state
\citep{OPAL}, OPAL opacities %
\citep{Igl96}, and the NACRE nuclear reaction rates \citep{NACRE}. Convection was treated according to the
mixing-length formalism \citep[MLT; ][]{bohm} and defined through the parameter $\alpha=\ell/H_p$, where $H_p$ is the pressure scale height.
The initial heavy-element mass fraction $Z_i$
was calculated from the iron abundance given in Table \ref{stellar_param} 
using the relation [Fe/H]$=\log(Z/X)-\log(Z/X)_{\odot}$, where $(Z/X)$ is the value at
the stellar surface. We obtained $Z/X=0.045\pm0.003$ assuming the 
solar value $(Z/X)_{\odot}=0.0245$ \citep{GN93}. 

The resulting evolutionary tracks are characterized by the input stellar mass $M$, the initial chemical
composition, and a mixing-length parameter.

The location of the star in the H-R diagram identifies $\tau$~Boo as being at the beginning of the 
main-sequence phase of core-hydrogen burning, with an internal content of hydrogen in the core of about $X_c=0.5$. As predicted by \citet{2009MNRAS.398.1383F}, it has a shallow convective region,   with a depth of about 
$D_{cz}\simeq0.1R$, typical of stars 
in this phase.
According to the stellar evolution constraints, given
the match with the   
observed atmospheric properties, and with the use of all the possible values of mass and metallicity,   
our computations show that the age of $\tau$ Boo is $0.9\pm0.5$~Gyr and the mass is $M=1.38\pm0.05 M_\odot$.
It is clear that the accuracy on the values of age and mass here inferred are dependent  
on the stellar model calculation procedure, and in particular on the physical and chemical inputs \citep[e.g.,][]{2014EAS....65...99L}. 
Here we simply note that lower values of the solar heavy-element abundance $(Z/X)_{\odot}$ led to a mass lower by 5\%
and an age higher by 30\%. These values agree with those deduced above within 2-$\sigma$ uncertainty limits.
The lithium line at $6708\AA$ could not be used to confirm the young age, since T$_{\rm eff}=6399 K$ (Table~\ref{stellar_param}) puts $\tau$~Boo in the lithium dip \citep[e.g.,][]{1995ApJ...446..203B}.

The value of age obtained by direct modeling is better constrained than the wide range of previous values obtained with different methods, such as the empirical relation between large-scale magnetic fluxes and age reported by \citet{vidotto}, chromospheric activity \citep{saffe,2000ApJ...531..415H}, isochrone techniques \citep{saffe,2001ApJ...549L.237S}, or X-ray luminosity \citep{sanzforcada}.

To reproduce the oscillations obtained in Sect.~\ref{sec:seismology}, 
we selected the models that best fit the observations among all the computed models.
For the selected models we calculated the
adiabatic oscillation frequencies using the ADIPLS code \citep{chris2008b}.
According to our calculations, this star should show solar-like pulsations with a spectrum
in the interval $700-2600~\mu$Hz and frequencies equally spaced by a large separation of 
about $\Delta \nu=    95\pm1~\mu$Hz. The observed large 
separation appears to agree excellently well with the theoretical separation. The
range of frequencies is very similar, although the spectral window
effects make a straight comparison very difficult. We conclude that there is no discrepancy between the values we inferred from the limited HARPS-N observations and the current
modeling of solar-like oscillations in $\tau$~Boo.

\subsection{Tidal evolution}\label{tidal}

\citet{2010A&A...512A..77L} and \citet{2014arXiv1411.3802D} found that the rotational period 
$P_{\rm rot}$ in a sample of main-sequence stars
with $T_{\rm eff} \ga 6300$ K accompanied by hot Jupiters with orbital 
period $P_{\rm orb}$ verifies the relationship
$1 \geq P_{\rm rot}/P_{\rm orb} \geq 2$ \citep[cf. Fig. 10 in][]{2014arXiv1411.3802D}.
Given its almost synchronous rotation, $\tau$ Boo satisfies this 
relationship as well.
The timescale of tidal synchronization of the stellar rotation is at 
least $1-2$ orders of magnitudes longer than the stellar age
\citep[e.g., ][]{2008MNRAS.385.1179D,2009MNRAS.395.2268B}, suggesting that the 
system reached the ZAMS close to its current synchronous state.
The subsequent evolution was characterized by a weak tidal interaction 
between the planet and the star with a very slow braking of the stellar 
rotation, owing to the low
braking efficiency of the stellar winds of mid-F type stars \citep[cf. ][Sect. 6.1.2 for details]{2014arXiv1411.3802D}.
In other words, the present orbital and 
rotational angular momenta in the $\tau$ Boo system are probably
close to the initial momenta with very little evolution since the system arrived on the ZAMS.


\section{Conclusions\label{sec:conclu}}

We studied the $\tau$~Boo system by means of
a new observational strategy applied to HARPS-N spectra that
were collected
in the framework of the GAPS project. This strategy allowed us to obtain
both high-cadence one-minute exposures to monitor the solar-like
oscillations, and high S/N unsaturated spectra to study stellar
activity and star-planet interaction.

We developed new computational tools to perform
the co-addition of consecutive spectra and the subsequent average of them,
preserving the correct computation of the BERV (and thus RV) values that have
to be referred to the flux-balanced reference time. Implementing a custom mask for $\tau$~Boo 
allowed us to extract more accurate RV values;
the adaptation of the Yabi platform gave us the possibility to test and verify these
new tools in a user-friendly system.

Our results are summarized in the following points.

\begin{itemize}

\item We updated the ephemeris for the planet $\tau$ Boo b and showed by means of our RVs that the binary companion $\tau$ Boo B is rapidly accelerating, confirming the astrometric predictions that it is approaching the periastron on a highly eccentric orbit. We plan to take additional observations in the next years to have a more reliable value of its eccentricity.

\item The SME software was applied to determine new stellar parameters, giving the same results as the equivalent width method. In particular, we refined the values of $\tau$~Boo's mass and radius to $1.39\pm0.25$ M$_{\odot}$ and $1.42\pm 0.08$ R$_{\odot}$.
We established that the star shows strong differential rotation, which we determined by means of both the LSD mean line profile and the CCF. We stress that the CCF computed by the HARPS-N pipeline can therefore be reliably used as an indicator of differential rotation.

\item 
The analysis of the correlations between several indices pointed out evident chromospheric activity.
In particular, the activity indicators extracted from HARPS-N spectra suggest the presence of a plage around one of the poles of the star.
The nature of the chromospheric activity remains uncertain. It is unclear if it is due to SPMI or to a corotating active region, or both.

\item Solar-like oscillations are detected in the RV time series. Although very limited by the spectral window, we inferred observational values of $\nu_{\rm max}$ and $\Delta \nu$. These values agree
well both with the scaling relations and the asteroseismic model computed from our stellar parameters.
This result supports our confidence in the application of the asteroseismic approach to
other bright stars hosting exoplanets to  
constrain their ages and masses.

\item From the evolutionary point of view, $\tau$~Boo is at the beginning of the main-sequence phase of core-hydrogen burning, with an age of $0.9\pm0.5$ Gyr. The model we built allowed us to further constrain the value of the stellar mass to $1.38\pm0.05$ M$_\odot$ and thus, using $i=44.5\pm1.5^\circ$ \citep{2012Natur.486..502B}, the mass of the planet to $6.13\pm 0.17$ M$_{\rm Jup}$.
\end{itemize}

\begin{acknowledgements}
{The GAPS project acknowledges
support from INAF through the ''Progetti Premiali'' funding scheme of
the Italian Ministry of Education, University, and Research. 
We thank the referee G. Walker for interesting and useful comments that helped improve the clarity of the paper.
}
\end{acknowledgements}


\begin{thebibliography}{}

\bibitem[Angulo et al.(1999)]{NACRE} Angulo, C., Arnould, M., Rayet M., et al. 1999, Nucl. Phys. A, 656, 3
\bibitem[Balachandran(1995)]{1995ApJ...446..203B} Balachandran, S.\ 1995, \apj, 446, 203 
\bibitem[Baliunas et al.(1997)]{1997ApJ...474L.119B} Baliunas, S.~L., Henry, G.~W., Donahue, R.~A., et al.\ 1997, \apjl, 474, L119 
\bibitem[Barker \& Ogilvie(2009)]{2009MNRAS.395.2268B} Barker, A.~J., \& Ogilvie, G.~I.\ 2009, \mnras, 395, 2268 
\bibitem[Bazot et al.(2012)]{18sco} Bazot, M., Ireland, M.J., Huber, D., et al. 2012, \aap,    
\bibitem[Bedding et al.(2007)]{betahyi} Bedding, T.R., Kjeldsen, H., Arentoft, T., et al. 2007, \apj, 663, 1315
\bibitem[Biazzo et al. (2012)]{2012MNRAS.427.2905B} Biazzo, K., D'Orazi, V., Desidera, S., et al.\ 2012, \mnras, 427, 2905 
\bibitem[B\"ohm-Vitense(1958)]{bohm} B\"ohm-Vitense, E., 1958, Zeitschrift f\"ur Astrophysik, 46, 1115
\bibitem[Borsa et al.(2013)]{internalsum} Borsa, F. , Rainer, M., Poretti, E. \ 2013, Internal report GAPS-SCI-REP-006
\bibitem[Brogi et al.(2012)]{2012Natur.486..502B} Brogi, M., Snellen, I.~A.~G., de Kok, R.~J., et al.\ 2012, \nat, 486, 502 
\bibitem[Butler et al.(1997)]{1997ApJ...474L.115B} Butler, R.~P., Marcy, G.~W., Williams, E., et al.\ 1997, \apjl, 474, L115 
\bibitem[Butler et al.(2006)]{2006ApJ...646..505B} Butler, R.~P., Wright, J.~T., Marcy, G.~W., et al.\ 2006, \apj, 646, 505 
\bibitem[Castelli \& Kurucz(2004)]{2004astro.ph..5087C} Castelli, F., \& Kurucz, R.~L.\ 2004, arXiv:astro-ph/0405087 
\bibitem[Catala et al.(2007)]{2007MNRAS.374L..42C} Catala, C., Donati, J.-F., Shkolnik, E., et al.\ 2007, \mnras, 374, L42 
\bibitem[Christensen-Dalsgaard(2008a)]{chris2008a} Christensen-Dalsgaard, J. 2008a, Ap\&SS, 316, 13
\bibitem[Christensen-Dalsgaard(2008b)]{chris2008b} Christensen-Dalsgaard, J. 2008b, Ap\&SS, 316, 113
\bibitem[Collier Cameron et al.(1999)]{1999Natur.402..751C} Collier Cameron, A., Horne, K., Penny, A., \& James, D.\ 1999, \nat, 402, 751 
\bibitem[Cosentino et al. (2012)]{2012SPIE.8446E..1VC} Cosentino, R., Lovis, C., Pepe, F., et al.\ 2012, \procspie, 8446,  
\bibitem[Covino et al.(2013)]{2013A&A...554A..28C} Covino, E., Esposito, M., Barbieri, M., et al.\ 2013, \aap, 554, A28 
\bibitem[Cunha et al.(2007)]{cunha} Cunha, M. S., Aerts C., Christensen-Dalsgaard J. et al. A\&ARv 14, 217
\bibitem[Damiani \& Lanza(2015)]{2014arXiv1411.3802D} Damiani, C., \& Lanza, A.~F., \ 2015, \aap, 574, A39
\bibitem[Di Mauro et al.(2011)]{dimauro2011} Di Mauro, M. P., Cardini, D., Catanzaro, G. et al., 2011, \mnras, 415, 3783
\bibitem[D{\'{\i}}az et al.(2007)]{2007MNRAS.378.1007D} D{\'{\i}}az, R.~F., Cincunegui, C., \& Mauas, P.~J.~D.\ 2007, \mnras, 378, 1007 
\bibitem[Donati et al.(2008)]{2008MNRAS.385.1179D} Donati, J.-F., Moutou, C., Far{\`e}s, R., et al.\ 2008, \mnras, 385, 1179 
\bibitem[Donati et al. (1997)]{1997MNRAS.291.658D} Donati, J.-F., Semel, M., Carter, B. D., et al.\ 1997, \mnras, 291, 658 
\bibitem[Drummond (2014)]{2014AJ....147...65D} Drummond, J.~D.\ 2014, \aj, 147, 65 
\bibitem[Dumusque et al.(2014)]{2014ApJ...796..132D} Dumusque, X., Boisse, I., \& Santos, N.~C.\ 2014, \apj, 796, 132 
\bibitem[Fares et al.(2009)]{2009MNRAS.398.1383F} Fares, R., Donati, J.-F., Moutou, C., et al.\ 2009, \mnras, 398, 1383 
\bibitem[Fares et al.(2013)]{Fares2013} Fares, R., Moutou, C., Donati, J.-F., et al.\ 2013, \mnras, 435, 1451 
\bibitem[Figueira et al.(2013)]{Figueira2013} Figueira, P., Santos, N.~C., Pepe, F., et al.\ 2013, \aap, 557, AA93 
\bibitem[Fischer et al.(2014)]{2014ApJS..210....5F} Fischer, D.~A., Marcy, G.~W., \& Spronck, J.~F.~P.\ 2014, \apjs, 210, 5
\bibitem[Flower (1996)]{1996ApJ...469..355F} Flower, P.~J.\ 1996, \apj, 469, 355  
\bibitem[Fortney et al.(2006)]{2006ApJ...642..495F} Fortney, J.~J., Saumon, D., Marley, M.~S., et al.\ 2006, \apj, 642, 495 
\bibitem[Garcia-Lopez et al.(1993)]{1993A&A...273..482G} Garcia-Lopez, R.~J., Rebolo, R., Beckman, J.~E., \& McKeith, C.~D.\ 1993, \aap, 273, 482 \bibitem[Gomes da Silva et al.(2014)]{2014A&A...566A..66G} Gomes da Silva, J., Santos, N.~C., Boisse, I., et al.\ 2014, \aap, 566, AA66 
\bibitem[Gratton(2013)]{internalgratton} Gratton, R. 2013, Internal report GAPS-SCI-REP-002
\bibitem[Gray(2008)]{Gray2008} Gray, D.~F.\ 2008, The Observation and Analysis of Stellar Photospheres, by David F.~Gray, Cambridge, UK: Cambridge University Press, 2008, 
\bibitem[Grevesse \& Noels(1993)]{GN93} Grevesse, N. \& Noels, A. 1993 in Origin and Evolution of the Elements, ed. S. Kubono \& T. Kajino, 14
\bibitem[Henry et al.(2000)]{2000ApJ...531..415H} Henry, G.~W., Baliunas, S.~L., Donahue, R.~A., et al.\ 2000, \apj, 531, 415 
\bibitem[Hunter et al.(2012)]{YABI} Hunter A.A., Macgregor A.B., Szabo T.O., et al., Yabi: An online research environment for Grid, High Performance and Cloud computing, Source Code for Biology and Medicine 2012, 7:1 
\bibitem[Iglesias \& Rogers(1996)]{Igl96} Iglesias, C. A., \& Rogers F. J. 1996, \apj, 464, 943
\bibitem[Landman(1981)]{1981ApJ...251..768L} Landman, D.~A.\ 1981, \apj, 251, 768
\bibitem[Lanza(2009)]{2009A&A...505..339L} Lanza, A.~F.\ 2009, \aap, 505, 339
\bibitem[Lanza(2010)]{2010A&A...512A..77L} Lanza, A.~F.\ 2010, \aap, 512, AA77 
\bibitem[Lanza(2012)]{2012A&A...544A..23L} Lanza, A.~F.\ 2012, \aap, 544, AA23
\bibitem[Lanza(2014)]{2014A&A...572L...6L} Lanza, A.~F.\ 2014, \aap, 572, LL6 
\bibitem[Lebreton et al.(2014)]{2014EAS....65...99L} Lebreton, Y., Goupil, M.~J., \& Montalb{\'a}n, J.\ 2014, EAS Publications Series, 65, 99 
\bibitem[Legendre \& Legendre(1983)]{Legendre1998} Legendre, P., Legendre, L., 1998, Numerical Ecology, Elsevier
\bibitem[Lockwood et al.(2014)]{2014ApJ...783L..29L} Lockwood, A.~C., Johnson, J.~A., Bender, C.~F., et al.\ 2014, \apjl, 783, L29 
\bibitem[Lovis et al.(2011)]{2011arXiv1107.5325L} Lovis, C., Dumusque, X., Santos, N.~C., et al.\ 2011, arXiv:1107.5325 
\bibitem[Mart{\'{\i}}nez-Arn{\'a}iz et al.(2010)]{2010A&A...520A..79M} Mart{\'{\i}}nez-Arn{\'a}iz, R., Maldonado, J., Montes, D., et al.\ 2010, \aap, 520, AA79 
\bibitem[Mart{\'{\i}}nez-Arn{\'a}iz et al.(2011)]{Martinez2011} Mart{\'{\i}}nez-Arn{\'a}iz, R., L{\'o}pez-Santiago, J., Crespo-Chac{\'o}n, I., \& Montes, D.\ 2011, \mnras, 414, 2629 
\bibitem[Mathur et al.(2014)]{Mathur2014} Mathur, S., Garc{\'{\i}}a, R.~A., Ballot, J., et al.\ 2014, \aap, 562, AA124 
\bibitem[Meunier \& Delfosse(2009)]{2009A&A...501.1103M} Meunier, N., \& Delfosse, X.\ 2009, \aap, 501, 1103 
\bibitem[Nardetto et al.(2006)]{Nardetto2006} Nardetto, N., Mourard, D., Kervella, P., et al.\ 2006, \aap, 453, 309 
\bibitem[Noyes et al.(1984)]{1984ApJ...279..763N} Noyes, R.~W., Hartmann, L.~W., Baliunas, S.~L., et al.\ 1984, \apj, 279, 763 
\bibitem[Pepe et al. (2002)]{2002A&A...388..632P} Pepe, F., Mayor, M., Galland, F., et al.\ 2002, \aap, 388, 632 
\bibitem[Queloz et al.(2001)]{2001A&A...379..279Q} Queloz, D., Henry, G.~W., Sivan, J.~P., et al.\ 2001, \aap, 379, 279 
\bibitem[Rainer(2013)]{internalmask} Rainer, M. 2013, Internal report GAPS-SCI-REP-007
\bibitem[Reiners \& Schmitt (2002)]{2002A&A...384..155R} Reiners, A., \& Schmitt, J.~H.~M.~M.\ 2002, \aap, 384, 155 
\bibitem[Reiners (2003)]{2003AA...408..707R} Reiners, A.\ 2003, A\&A, 408, 707
\bibitem[Reiners \& Schmitt (2003)]{2003A&A...398..647R} Reiners, A., Schmitt, J.H.M.M.,\ 2003, A\&A, 398, 647
\bibitem[Reiners (2006)]{2006A&A...446..267R} Reiners, A.\ 2006, A\&A, 446, 267
\bibitem[Roberts et al. (2011)]{2011AJ....142..175R} Roberts, L.~C., Jr., Turner, N.~H., ten Brummelaar, T.~A., et al.\ 2011, \aj, 142, 175 
\bibitem[Rodler et al.(2010)]{2010A&A...514A..23R} Rodler, F., K{\"u}rster, M., \& Henning, T.\ 2010, \aap, 514, A23 
\bibitem[Rodler et al.(2012)]{2012ApJ...753L..25R} Rodler, F., Lopez-Morales, M., \& Ribas, I.\ 2012, \apjl, 753, L25 
\bibitem[Rogers \& Nayvonov(2002)]{OPAL} Rogers, F. J., Nayvonov, A. 2002, \apj, 576, 1064
\bibitem[Saffe et al.(2005)]{saffe} Saffe, C., G{\'o}mez, M., \& Chavero, C.\ 2005, \aap, 443, 609 
\bibitem[Santos et al.(2004)]{santos} Santos, N.C., Israelian, G., Mayor, M., et al. 2004, \aap, 415, 1153 
\bibitem[Santos et al.(2013)]{2013A&A...556A.150S} Santos, N.~C., Sousa, S.~G., Mortier, A., et al.\ 2013, \aap, 556, AA150 
\bibitem[Sanz-Forcada et al.(2010)]{sanzforcada} Sanz-Forcada, J., Ribas, I., Micela, G., et al.\ 2010, \aap, 511, LL8 
\bibitem[Scandariato et al.(2013)]{Scandariato2013} Scandariato, G., Maggio, A., Lanza, A.~F., et al.\ 2013, \aap, 552, AA7 
\bibitem[Shkolnik et al.(2005)]{2005ApJ...622.1075S} Shkolnik, E., Walker, G.~A.~H., Bohlender, D.~A., et al. \ 2005, \apj, 622, 1075 
\bibitem[Shkolnik et al.(2008)]{2008ApJ...676..628S} Shkolnik, E., Bohlender, D.~A., Walker, G.~A.~H., \& Collier Cameron, A.\ 2008, \apj, 676, 628 
\bibitem[Sousa et al. (2007)]{2007A&A...469..783S} Sousa, S.~G., Santos, N.~C., Israelian, G., et al.\ 2007, \aap, 469, 783 
\bibitem[Stello et al.(2009)]{stello} Stello, D., Chaplin, W.J., Basu, S., et al. 2009, \mnras, 400, L80 
\bibitem[Stelzer et al.(2013)]{2013A&A...558A.141S} Stelzer, B., Frasca, A., Alcal{\'a}, J.~M., et al.\ 2013, \aap, 558, AA141 
\bibitem[Suchkov \& Schultz(2001)]{2001ApJ...549L.237S} Suchkov, A.~A., \& Schultz, A.~B.\ 2001, \apjl, 549, L237
\bibitem[Tody(1993)]{Tody1993} Tody, D.\ 1993, Astronomical Data Analysis Software and Systems II, 52, 173 
\bibitem[Torres (2010)]{2010AJ....140.1158T} Torres, G.\ 2010, \aj, 140, 1158 
\bibitem[Tripicchio et al.(1997)]{Tripicchio97} Tripicchio, A., Severino, G., Covino, E., et al.\ 1997, \aap, 327, 681 
\bibitem[Valenti \& Piskunov (1996)]{1996AA.118.595} Valenti, J.A., Piskunov. N.E.\ 1996, A\&A, 118, 595
\bibitem[van Leeuwen (2007)]{2007A&A...474..653V} van Leeuwen, F.\ 2007, \aap, 474, 653 
\bibitem[Vani\^cek(1971)]{vani} Vani\^cek, P. 1971, \apss, 12, 10
\bibitem[Vidotto et al.(2014)]{vidotto}Vidotto, A.A.,  Gregory, S.G.,  Jardine, M. et al., 2014, \mnras, 441, 2361
\bibitem[Walker et al.(2008)]{2008A&A...482..691W} Walker, G.~A.~H., Croll, B., Matthews, J.~M., et al.\ 2008, \aap, 482, 691 
\bibitem[Wang et al.(2012)]{2012ApJ...761...46W} Wang, X., Sharon, Wright, J.~T., Cochran, W., et al.\ 2012, \apj, 761, 46 
\bibitem[Wright \& Howard(2009)]{2009ApJS..182..205W} Wright, J.~T., \& Howard, A.~W.\ 2009, \apjs, 182, 205 
\bibitem[Zechmeister \& Kurster(2009)]{scargle} Zechmeister, M., \& K\"urster, M. 2009, \aap, 496, 577

\end{thebibliography}
\end{document}